\documentclass[prb,amsmath,amssymb,twocolumn,showpacs]{revtex4}
\usepackage{graphicx}
\usepackage{subfigure}
\usepackage{bm}
\usepackage{color}

\DeclareMathAlphabet{\mathpzc}{OT1}{pzc}{m}{it} 
\begin{document}
\title{Tight-binding theory of spin-orbit coupling in graphynes}
\author{Guido van Miert, Vladimir Juri\v ci\' c, and Cristiane Morais Smith}

\affiliation{Institute for Theoretical Physics, Centre for Extreme Matter and Emergent Phenomena, Utrecht University, Leuvenlaan 4, 3584 CE Utrecht, The Netherlands}

\begin{abstract}
We investigate the effects of Rashba and intrinsic spin-orbit couplings (SOC) in graphynes. First, we develop a general method to address spin-orbit couplings within the tight-binding theory. Then, we apply this method to $\alpha$-, $\beta$-, and $\gamma$-graphyne, and determine the SOC parameters in terms of the microscopic hopping and on-site energies. We find that for $\alpha$-graphyne, as in graphene, the intrinsic SOC opens a non-trivial gap, whereas the Rashba SOC splits each Dirac cone into four. In $\beta$- and $\gamma$-graphyne, the Rashba SOC can lead to a Lifshitz phase transition, thus transforming the zero-gap semiconductor into a gapped system or vice versa, when pairs of Dirac cones annihilate or emerge. The existence of internal (within the benzene ring) and external SOC in these compounds allows us to explore a myriad of phases not available in graphene.
\end{abstract} 
\pacs{73.22.-f, 81.05.Zx, 31.15.aj, 31.15.ae}
\maketitle
\section{Introduction}
During the last decade, graphene has attracted enormous attention, and has provided a new paradigm for studying pseudo-relativistic fermions in  condensed-matter systems [\onlinecite{RMP-CastroNeto2009}]. The peculiar Dirac-type structure
of its low-energy quasiparticles arises due to the lattice geometry and time-reversal symmetry. The honeycomb lattice, which consists of two equivalent interpenetrating triangular lattices, gives rise to the touching of the valence and conduction bands at two inequivalent $K$ and $K'$ points at the corners of the hexagonal Brillouin zone (BZ), which are related by time-reversal symmetry. Although the first proposal for a time-reversal topological insulator invoked graphene [\onlinecite{Kane-Mele-PRL2005}], its experimental realization has been hampered by the weak spin-orbit coupling (SOC) in this material. On the other hand, this created a lot of activity towards the tailoring of artificial structures exhibiting  Dirac cones and strong SOC. Some of the so far proposed systems include self-assembled honeycomb arrays of CdSe and PbSe semiconducting nanocrystals [\onlinecite{MoraisSmith-PRX2014}], patterned quantum dots [\onlinecite{Pellegrini}], and molecular graphene [\onlinecite{manoharan}]. Yet another interesting class of Dirac materials in this respect consists of graphynes.

Graphynes are two-dimensional carbon allotropes that differ from graphene by the presence of triple bonds ($-$C $\equiv$ C$-$) into their lattice structure [\onlinecite{graphyne-review2014}]. Figure~\ref{fig:lattices} displays the lattice structure of $\alpha$-, $\beta$-, and $\gamma$-graphyne. They have not been experimentally realized yet, as opposed to graphdiyne [\onlinecite{Li2010}], which features {\it pairs} of acetylene bonds in its crystal lattice. Since their proposal in 1987 [\onlinecite{baughman1987}], they have attracted considerable interest, especially because of their band structure, which exhibits Dirac-like properties [\onlinecite{narita1998,tahara2007,kang2011,yue2012,cranford2011,peeters2013}]. In particular, it has been shown by using {\it ab initio} and tight-binding (TB) methods [\onlinecite{malko2012}] that $\alpha$-graphyne features Dirac cones at the high-symmetry $K$ and $K'$ points of the BZ, whereas in $\beta$-graphyne they occur along the high-symmetry $\Gamma-M$ line. On the other hand, $\gamma$-graphyne is gapped. A criterion for the existence of the Dirac cones has been provided within simple TB models [\onlinecite{Huang2013}]. Furthermore, the possibility of manipulation of the Dirac cones by chemical reactions has been discussed [\onlinecite{JJZheng2013}]. The control of the electronic properties with adatoms has been considered in Refs.~[\onlinecite{Kim-Choi2012, zhengJChemPhys2013, ZhangJPhysChemC2011, HwangJPhysChemC2012, HeJPhysChemC2012}], which  is particularly important in light of inducing topological  properties in graphyne-based materials.

Topologically nontrivial properties of the electronic band structure of a material, from a practical perspective, critically depend on the strength of the spin-orbit interaction. On the other hand, graphynes are based on carbon and, as such, are expected to feature a weak SOC, as it is the case in graphene, for instance. The effect of intrinsic SOC in these systems has been recently investigated using {\it ab initio} methods [\onlinecite{MZhaoSciRep2013}]. Furthermore, the above-mentioned possibility of controlling the electronic properties of graphynes with adatoms puts forward a way of manipulating SOC in these systems by using adatoms of heavy elements, such as Bi and Sn, for instance. Before doing so, however, a general framework for addressing SOC in graphynes has to be developed. This is precisely the aim of the present paper. The effect of SOC in $\beta$-graphyne has been previously investigated by the same authors [\onlinecite{MMJ}].
Here, we derive a general TB theory of the spin-orbit interactions in graphynes, apply it to $\alpha$- and $\gamma$-graphynes, as well as provide a comparison of the effects of the SOC in the three compounds. For completeness we also repeat some of our results on $\beta-$graphyne previously reported in Ref.\ [\onlinecite{MMJ}].

We  concentrate on the effect of both Rashba SOC, which can be induced by an external electric field, as well as on the intrinsic SOC. We find that the spin-orbit interactions produce different effects for $\alpha$-, $\beta$-, and $\gamma$-graphyne. In $\alpha$- and $\beta$-graphyne [\onlinecite{MMJ}], the intrinsic SOC opens up a non-trivial band gap. The Rashba SOC affects $\alpha$-graphyne in exactly the same way as it does for graphene: it lifts the spin degeneracy and splits each Dirac cone into four distinct Dirac cones. In $\beta$-graphyne the Rashba SOC splits each Dirac cone into two, instead of four. As the coupling is increased, this pair of cones eventually merges with another pair on the line connecting the $K$ and $K'$ points in the Brillouin zone. When the coupling is further increased, a new pair of Dirac cones emerges at the line connecting the $\Gamma$ and $M$ points. Finally, in $\gamma$-graphyne the effect is the opposite as compared to $\beta$-graphyne, since now one begins with a gapped system, and if the Rashba SOC parameter exceeds a certain value, the Dirac cones emerge along the line connecting the $K$ and $K'$ points.

In the following, we first introduce in Sec.~II the TB model for the different types of graphyne  and derive the corresponding band structures without the SOC. Then, we investigate the form of the Rashba and the intrinsic SOC in Sec.~III and introduce an effective model in Sec. IV. Our conclusions are drawn in Sec.~V, and the details of the calculations are presented in appendices.
\section{Tight-binding model}\label{sec2}
\begin{figure*}[t]
\centering

\includegraphics[width=\textwidth]{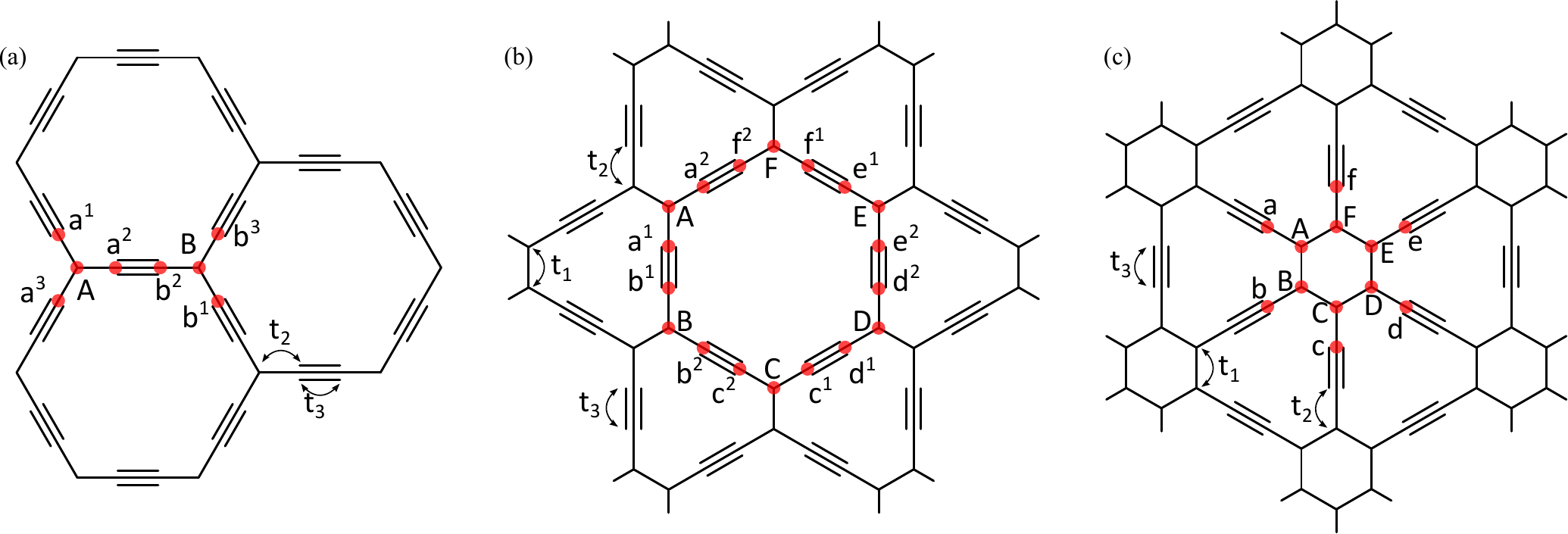}

\caption{\label{fig:lattices} (Color online) Lattice structure of $\alpha$-, $\beta$- and $\gamma$-graphyne, shown in panels (a), (b), and (c), respectively. Atoms at the vertices are denoted by capital letters, whereas atoms located at the edges are denoted by lower-case letters.  The hopping parameters $t_i$ are also shown, with the subscripts $1$, $2$, and $3$ corresponding to vertex-vertex,  vertex-edge, and edge-edge hoppings, respectively.}
\label{fig:lattices}
\end{figure*}
The three types of graphyne can all be described in terms of a TB model that takes into account only the $p_z$ orbitals. It was also shown that one can integrate out the contributions coming from the acetylene bonds to derive an effective model (see also Appendix~\ref{eff})[\onlinecite{zhe,Kim-Choi2012}]. By doing so, one can describe $\alpha$-graphyne  by the same Hamiltonian as used for graphene, but with a different value of the nearest-neighbor (NN) hopping parameter. On the other hand, $\beta$- and $\gamma$-graphyne are well described by an effective six-site model, with two hopping parameters. In the following, we discuss separately the band structure for each type of graphyne and describe the system by an effective Hamiltonian.

\subsection{$\alpha$-graphyne}
Of all graphynes, $\alpha$-graphyne is the simplest. One can envision $\alpha$-graphyne as being obtained from graphene upon insertion of two more carbon atoms bonded by an acetylene linkage between any two carbon atoms of the honeycomb graphene lattice. As a result, the number of atoms in the unit cell grows from $2$ to $8$ [see Fig.~\ref{fig:lattices}(a)].  To describe this system in terms of a TB model, we will need two different hopping parameters: $t_{\alpha,2}$ and $t_{\alpha,3}$. Using the labeling shown in Fig.~\ref{fig:lattices}(a), the Hamiltonian reads as
\begin{align}\label{eqha}
H^\alpha&=t_{\alpha,2}\sum_{\langle i,j\rangle}\left[A^\dagger_i\left(a_{1,j}+a_{2,j}+a_{3,j}\right)+B^\dagger_i\left(b_{1,j}\right.\right.\\
&+\left.\left. b_{2,j}+b_{3,j}\right)\vphantom{A^\dagger_i}\right]+t_{\alpha,3}\sum_{\langle i,j\rangle}\left(a^\dagger_{1,i}b_{1,j}+a^\dagger_{2,i}b_{2,j}\right.\nonumber\\
&+\left.a^\dagger_{3,i}b_{3,j}\right)+h.c.\nonumber
\end{align}

By integrating out the electrons forming the acetylene bonds (see Appendix~\ref{appaa}), we obtain an effective low-energy Hamiltonian
\begin{align}\label{aeff}
H^\alpha_{\rm{eff}}&=\tilde{t}_\alpha\sum_{\langle i,j\rangle}A^\dagger_i B_j+h.c.,
\end{align}
where $\tilde{t}_\alpha=-t_{\alpha,2}^2 t_{\alpha,3}/(3 t_{\alpha,2}^2+t_{\alpha,3}^2)$. Fitting the TB-parameters with a first-principles calculation [\onlinecite{zhe}] yields $t_{\alpha,2}=-2.85$eV and $t_{\alpha,3}=-7.50$eV, hence $\tilde{t}_{\alpha}=0.76$eV.  As shown in Fig.~\ref{fig:dispersion}(a), the band  structure obtained from the low-energy approximation (red dashed lines) agrees very well with the band structure obtained from the full TB model (blue solid lines).

Since the physics around the Fermi energy in $\alpha$-graphyne is described by the same Hamiltonian as graphene, it comes as no surprise that $\alpha$-graphyne exhibits two Dirac cones at the $K$ and $K'$ points, see Figs.~\ref{fig:dispersion}(a) and \ref{fig:dispersion2}(a). The main difference with graphene is the reduced Fermi velocity. In graphene the Fermi velocity is given by $v_F= 3 a t/2\hbar\simeq10^6$ m/s, with $t\approx -2.8$eV the hopping amplitude [\onlinecite{RMP-CastroNeto2009}] and $a$ the NN distance, whereas in $\alpha$-graphyne $v_F= 9 a \tilde{t}_\alpha/2\hbar\simeq7\times10^5$ m/s. Note that in graphene the NN distance $a=1.42$\AA [\onlinecite{RMP-CastroNeto2009}], whereas in $\alpha$-graphyne there are actually  two different bond lengths, one for the single bond $d_s=1.40$\AA~ and one for the triple bond $d_t=1.23$\AA [\onlinecite{Kim-Choi2012}]. However, setting $d_s$ and $d_t$ equal to the bond length $a$ in graphene yields an error smaller than $10\%$. As a consequence of the reduced Fermi velocity, many-body effects arising from the long-range Coulomb interaction, with effective coupling constant $\alpha=e^2/v_F$, where $e$ is the electron charge, could be more pronounced in graphyne than in graphene.
\subsection{$\beta$-graphyne}
Among the three different types of graphyne that we consider, $\beta$-graphyne has the most complicated lattice structure. Its unit cell involves $18$ atoms and consists of a hexagon, which has one carbon atom located at each vertex, and two carbon atoms connected by an acetylene bond between each two neighboring vertices [see Fig.~\ref{fig:lattices}(b)]. A TB description of $\beta$-graphyne requires three different hopping parameters: $t_{\beta,1}$, $t_{\beta,2}$, and $t_{\beta,3}$. Using the labeling displayed in Fig. \ref{fig:lattices}(b), the TB Hamiltonian reads as
\begin{align}\label{eqhb}
H^\beta &=t_{\beta,1}\sum_{\langle i,j\rangle} \left(A^\dagger_i D_j+B^\dagger_i E_j+C^\dagger_i F_j\right)\\
&+t_{\beta,2}\sum_{\langle i,j\rangle}\left[A^\dagger_i\left(a_{1,j}+a_{2,j}\right)+B^\dagger_i\left(b_{1,j}+b_{2,j}\right)\right.\nonumber\\
&+C^\dagger_i\left(c_{1,j}+c_{2,j}\right)+D^\dagger_i\left(d_{1,j}+d_{2,j}\right)+E^\dagger_i\left(e_{1,j}+e_{2,j}\right)\nonumber\\
&\left.+F^\dagger_i\left(f_{1,j}+f_{2,j}\right)\vphantom{A^\dagger_i}\right]+t_{\beta,3}\sum_{\langle i,j\rangle}\left(a^\dagger_{1,i}b_{1,j}+b^\dagger_{2,i}c_{2,j}\right.\nonumber\\
&\left.+c^\dagger_{1,i}d_{1,j}+d^\dagger_{2,i}e_{2,j}+e^\dagger_{1,i}f_{1,j}+f^\dagger_{2,i}a_{2,j}\right)+h.c.\nonumber
\end{align}

By performing a Fourier transformation and subsequently eliminating the high-energy orbitals (see Appendix~\ref{appb}), we obtain an effective six-site model. The effective low-energy Hamiltonian reads as
\begin{align}\label{effb}
H^\beta_{\rm{eff}}&=t^{\beta}_{\rm int}\sum_{\langle i,j\rangle}\left[A^\dagger_i\left(B_j+F_j\right)+C^\dagger_i\left(B_j+D_j\right)+E^\dagger_i\left(D_j\right.\right.\nonumber\\
&\left.\left.+F_j\right)\right]+t^{\beta}_{\rm ext}\sum_{\langle i,j\rangle}\left[A^\dagger_i D_j+C^\dagger_i F_j+E^\dagger_i B_j\right]+h.c.,
\end{align}
where $t^{\beta}_{\rm int}=-t_{\beta,2}^2 t_{\beta,3}/(2t_{\beta,2}^2+t_{\beta,3}^2)$ and $t^{\beta}_{\rm ext}=t_{\beta,1} t_{\beta,3}^2/(2t_{\beta,2}^2+t_{\beta,3}^2)$. In $\beta$-graphyne, it is found [\onlinecite{zhe}] that $t_{\beta,1}=-2.00$eV, $t_{\beta,2}=-2.70$eV, and $t_{\beta,3}=-4.30$eV, hence $t^{\beta}_{\rm int}=0.95$eV and $t^{\beta}_{\rm ext}=-1.12$eV. It turns out that the agreement between this effective model (red dashed lines) and the full TB Hamiltonian (blue solid lines) is extremely good [see Fig.~\ref{fig:dispersion}(b)].
The dispersion relation exhibits six Dirac cones, located on the line $\Gamma-M$ [see also Fig.~\ref{fig:dispersion2}(b)]. As opposed to graphene and $\alpha$-graphyne, where the cones exhibit a threefold  symmetry, in $\beta$-graphyne the cones are symmetric under mirror reflection through the normal plane containing the line $\Gamma$-$M$ [\onlinecite{malko2012}].

\subsection{$\gamma$-graphyne}
\begin{figure*}[t]
\centering
\includegraphics[width=\textwidth]{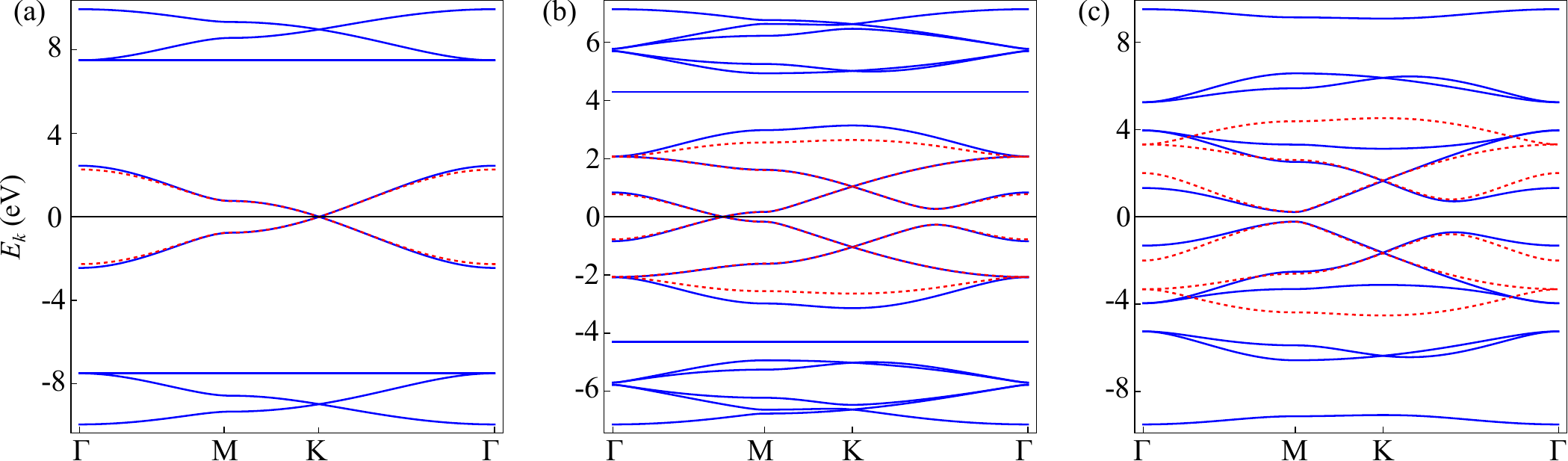}
\caption{\label{fig:dispersion} (Color online) Dispersion relation for $\alpha$-, $\beta$-, and $\gamma$-graphyne along high-symmetry lines, shown in (a), (b), and (c), respectively. The (blue) solid lines correspond to the dispersion relation obtained from the full TB Hamiltonian, whereas the (red) dashed lines correspond to the dispersion relation obtained from the low-energy approximation.}
\end{figure*}

\begin{figure}[b]
\centering
\subfigure[]{
\includegraphics[width=0.14\textwidth]{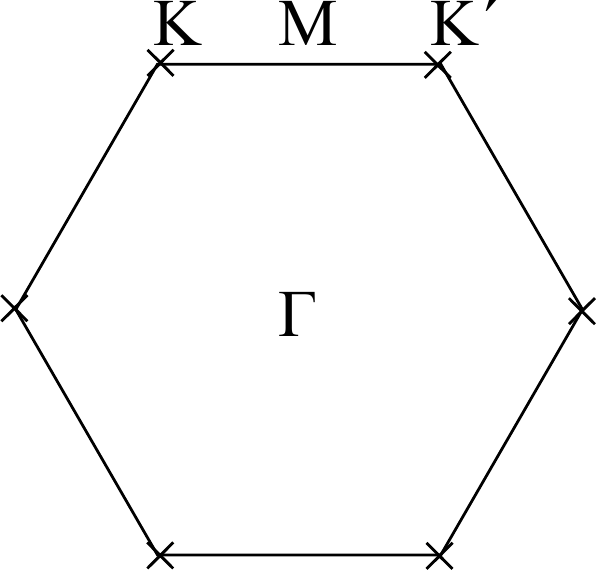}}~
\subfigure[]{
\includegraphics[width=0.14\textwidth]{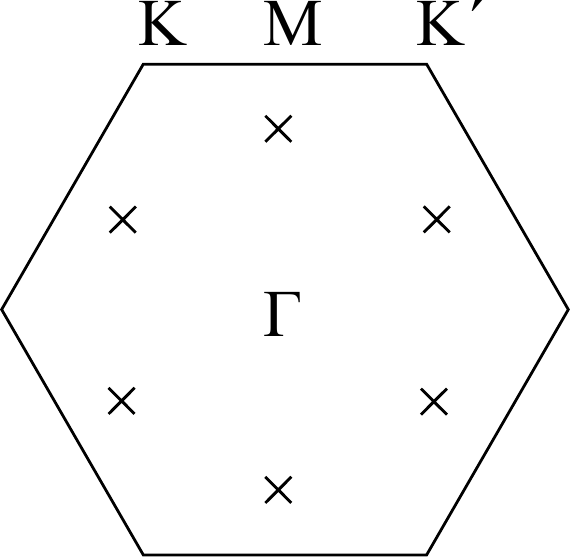}}~
\subfigure[]{
\includegraphics[width=0.14\textwidth]{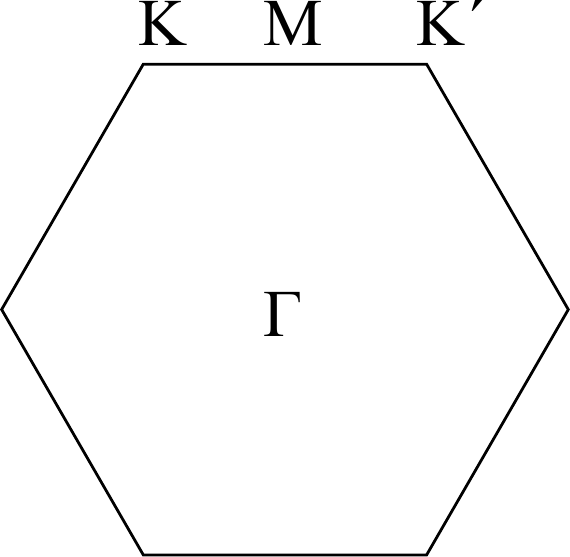}}
\caption{\label{fig:dispersion2} Sketch of the BZ for $\alpha$-, $\beta$-, and $\gamma$-graphyne, shown in panels (a), (b), and (c), respectively. The crosses correspond to Dirac cones.}
\end{figure}

$\gamma$-graphyne has a somewhat simpler structure than $\beta$-graphyne, as its unit cell contains only $12$ atoms [see Fig.~\ref{fig:lattices}(c)]. The TB description of $\gamma$-graphyne  involves three hopping parameters: $t_{\gamma,1}$, $t_{\gamma,2}$, and $t_{\gamma,3}$. Using the labeling shown in Fig.~\ref{fig:lattices}(c), we find that $H^\gamma$ is given by
\begin{align}\label{eqhc}
H^{\gamma}&=t_{\gamma,1}\sum_{\langle i,j\rangle}\left[A^\dagger_i\left(B_j+F_j\right)+C^\dagger_i\left(B_j+D_j\right)\right.\\
&\left.+ E^\dagger_i\left(D_j+F_j\right)\vphantom{A^\dagger_i}\right]+t_{\gamma,2}\sum_{\langle i,j\rangle}\left(A^\dagger_i a_j+B^\dagger_i b_j+C^\dagger_ic_j\right.\nonumber\\
&\left.+D^\dagger_id_j+E^\dagger_ie_j+F^\dagger_if_j\right)+t_{\gamma,3}\sum_{\langle i,j\rangle}\left(a^\dagger_id_j+b^\dagger_ie_j\right.\nonumber\\
&\left.+c^\dagger_if_j\right)+h.c.\nonumber
\end{align}
As for $\beta$-graphyne, here we also perform a Fourier transformation (see Appendix~\ref{appc}), and then eliminate the high-energy orbitals, to obtain the effective model
\begin{align}\label{effc}
H^\gamma_{\rm{eff}}&=t^{\gamma}_{\rm int}\sum_{\langle i,j\rangle}\left[A^\dagger_i\left(B_j+F_j\right)+C^\dagger_i\left(B_j+D_j\right)+E^\dagger_i\left(D_j\right.\right.\nonumber\\
&\left.\left.+F_j\right)\right]+t^{\gamma}_{\rm ext}\sum_{\langle i,j\rangle}\left[A^\dagger_i D_j+C^\dagger_i F_j+E^\dagger_i B_j\right]+h.c.,
\end{align}
with $t^{\gamma}_{\rm int}=t_{\gamma,1} t_{\gamma,3}^2/(t_{\gamma,2}^2+t_{\gamma,3}^2)$ and $t^{\gamma}_{\rm ext}=-t_{\gamma,2}^2 t_{\gamma,3}/(t_{\gamma,2}^2+t_{\gamma,3}^2)$. Mapping this TB model to DFT calculations [\onlinecite{zhe}] yields  $t_{\gamma,1}=-2.75$eV, $t_{\gamma,2}=-3.11$eV, and $t_{\gamma,3}=-4.04$eV, hence $t^{\gamma}_{\rm int}=-1.73$eV and $t^{\gamma}_{\rm ext}=1.50$eV. The band structure does not exhibit any Dirac points at the Fermi energy [see Figs.~\ref{fig:dispersion}(c) and \ref{fig:dispersion2}(c)]. The gap at the $M$ point is approximately equal to $0.44$eV. Notice that the low-energy approximation is less accurate for $\gamma$-graphyne than for $\alpha$- and $\beta$-graphyne due to the presence of a band gap.

\section{Spin-orbit coupling}
Whereas the SOC has been extensively studied in graphene, this coupling has so far been rather unexplored in graphynes. In this section, we derive the TB Hamiltonians corresponding to both Rashba and intrinsic SOC. Moreover, we show how the effective coupling parameters are related to the microscopic hopping parameters.\\

The intrinsic SOC originates from relativistic corrections to the Schr\"odinger equation. By expanding the Dirac equation up to second order in $v/c$, with $v$ denoting the electron velocity, one finds that the microscopic Hamiltonian acquires an additional term
\begin{align}
H_{L}&=-\frac{\hbar}{4mc^2}{\bm\sigma}\cdot\left( {\bf p}\times\nabla V\right),
\end{align}
where ${\bm\sigma}$ is a vector of Pauli matrices, ${\bf p}$ is the momentum, $m$ is the electron mass, and $V$ is the nuclear potential [\onlinecite{sakurai}]. If one rewrites this expression in spherical coordinates, one obtains
\begin{align}
H_{L}&=-f(r){\bm\sigma}\cdot {\bf L},
\end{align}
where $f$ is a function that goes rapidly to zero away from the origin and ${\bf L}$ is the orbital angular momentum. In order to see how this additional Hamiltonian enters the TB models, we need to reconsider the derivation presented in Sec.~\ref{sec2}. The models  we considered therein only describe the band-structure due to the $p_z$ orbitals, but the bonds in graphyne are formed by the $s$, $p_x$, and $p_y$ orbitals, called $\sigma$ orbitals due to their symmetry. When SOC is not involved, the $\sigma $ and $p_z$ orbitals decouple since the former are even and the latter are odd with respect to mirror reflection through the $x-y$ plane.  The inclusion of spin changes this picture drastically. Reflection through the $x-y$ plane is represented by $\sigma_z$ in spin-space and therefore spin up ($|\uparrow\rangle$) is even, whereas spin down ($|\downarrow\rangle$) is odd under this transformation.  Hence, this symmetry allows now for the coupling between $p_{z,\uparrow}$, $p_{x,\downarrow}$, $p_{y,\downarrow}$, and $s_{\downarrow}$ orbitals, which are odd under this reflection. Analogously, it follows that the orbitals $p_{z,\downarrow}$, $p_{x,\uparrow}$, $p_{y,\uparrow}$,  and $s_{\uparrow}$,  even under this symmetry operation, can be coupled. Moreover, when an external electric field is applied perpendicularly to the $x-y$ plane, the microscopic Hamiltonian includes an extra term
\begin{align}
 H_{E}&=E z,
\end{align}
with $E$ the magnitude of the applied electric field. This microscopic Hamiltonian $H_E$ couples now the $s$ orbitals to the $p_z$ orbitals. This occurs as a consequence of the broken mirror symmetry. It turns out that the combination of the terms $H_L$ and $H_E$ leads to the Rashba SOC, while $H_L$ alone leads to the intrinsic SOC. Hence, to describe SOC in graphyne it is necessary to include both the $p_z$- and the $\sigma$-orbitals. On top of the SOC generated by the $\sigma$-orbitals, we also need to consider the effect  of the $d_{xz}$ and $d_{yz}$ orbitals, as it has been shown before for graphene [\onlinecite{fabian}]. However, the nature of this effect is entirely different in graphynes, because even without considering spin, the $p_z$ orbitals already couple to the $d_{xz}$ and $d_{yz}$ orbitals. The intrinsic SOC leads to a spin dependent on-site hoppings between the $d_{xz}$ and $d_{yz}$ orbitals. In the following, we discuss separately the SOC generated by the $\sigma$-orbitals and by the $d$-orbitals.
\subsection{Spin-orbit coupling generated by the $\sigma$-orbitals}
Since we have to include the $\sigma$-orbitals and spin, the number of orbitals in the TB models increases by a factor of 8 (2 for spin, 4 for orbitals). The corresponding Hamiltonian reads as
\begin{align}
H&=H_{z}+H_{\sigma}+H^{z,\sigma}_{SOC}+\left(H^{z,\sigma}_{SOC}\right)^\dagger,
\end{align}
where $H_{z}$  describes the $p_z$ orbitals, $H_\sigma$ describes the $\sigma$ orbitals (see Appendix~\ref{sigmatb}), and $H^{z,\sigma}_{SOC}$  accounts for the hopping from $p_z$ orbitals to $\sigma$-orbitals due to the SOC. The latter can be decomposed as
\begin{align}
H^{z,\sigma}_{SOC}&=H^{z,\sigma}_{L}+H^{z,\sigma}_{E},
\end{align}
with the orbital angular momentum and electric field terms given by
\begin{align}\label{soch}
H^{z,\sigma}_{L}&=\xi_{p1}\sum_{i}{}^{'}p^\dagger_{z,i}\left(-i\sigma_y p_{x,i}+i\sigma_x p_{y,i}\right)\\
&+\xi_{p2}\sum_{i}{}^{''}p^\dagger_{z,i}\left(-i\sigma_y p_{x,i}+i\sigma_x p_{y,i}\right),\nonumber\\\label{soch2}
H^{z,\sigma}_{E}&=\xi_{sp1}\sum_{i}{}^{'}p^\dagger_{z,i}s_{i}+\xi_{sp2}\sum_{i}{}^{''}p^\dagger_{z,i}s_{i},
\end{align}
where $p_{z,i}^\dagger$ creates an electron in a $p_z$ orbital at position $i$, and analogous notation is used for the $p_x$, $p_y$, and $s$ orbitals. The exact value of the on-site coupling parameters $\xi_{p1}$, $\xi_{p2}$, $\xi_{sp1}$, and $\xi_{sp2}$ may be obtained by fitting the band structure to first-principles calculations. Note that $\xi_{sp1}$ and $\xi_{sp2}$ are both linear in $E$. Furthermore, the prime (double prime) in the summation indicates that the sum is taken over atoms located at the  edges (vertices). These terms result from considering the matrix elements ${\bm \sigma}\cdot{\bf L}$ and $E z$ (see also Table~\ref{tablematrix}). Let us consider the matrix element $\langle p_z|{\bm \sigma}\cdot{\bf L}|p_x\rangle$, as an example. As a first step, we rewrite $L_x$ and $L_y$  in terms of raising and lowering operators,
 \begin{align}
L_x&=\frac{1}{2}(L_++L_-),\nonumber\\
L_y&=-\frac{i}{2}(L_+-L_-).
 \end{align}
 Next, one rewrites the atomic orbitals $|p_x\rangle$, $|p_y\rangle$, and $|p_z\rangle$ in terms of the simultaneous eigenstates of the operators $H$, $L^2$ and $L_z$, $|n,l,m\rangle$. Since all the orbitals we consider are in the second shell, $n=2$,  we simply write $|2,l,m\rangle\equiv|l,m\rangle$. Then, we have
 \begin{align}
 |p_x\rangle&=\frac{1}{\sqrt{2}}(-|1,1\rangle+|1,-1\rangle),\nonumber\\
 |p_y\rangle&=\frac{i}{\sqrt{2}}(|1,1\rangle+|1,-1\rangle),\nonumber\\
 |p_z\rangle&=|1,0\rangle,
 \end{align}
 yielding
 \begin{align}
 L_x |p_x\rangle&=\frac{1}{2}\frac{1}{\sqrt{2}}(L_++L_-)(-|1,1\rangle+|1,-1\rangle)\nonumber\\
 &=\frac{1}{2}(-|1,0\rangle+|1,0\rangle)=0.
 \end{align}
 Similarly, we find
\begin{equation}
L_y| p_x\rangle=-i| p_z\rangle.
\end{equation}
As a result, we obtain $\langle p_z|{\bm \sigma}\cdot{\bf L}|p_x\rangle=\langle p_z|-i\sigma_y|p_z\rangle=-i\sigma_y$,  where we also used $\langle p_z|L_z|p_x\rangle=0$.

\begin{table}[b]
\centering
\begin{tabular}{ l || c | c | c  }
   ${\bm\sigma}\cdot {\bm L}$& $p_x$ & $p_y$ & $s$ \\
  \hline\hline
$p_z$& $-i\sigma_y$  & $i\sigma_x$ & $0$
\end{tabular}
\caption{Matrix elements for ${\bm\sigma}\cdot{\bf L}$. Note that the Pauli matrices act in spin space.}
\label{tablematrix}
\end{table}
Since the $p_z$-orbitals correspond to the low-energy states, we can use the approximation scheme outlined in the App.~\ref{eff}, which yields
\begin{align}\label{eq14}
H^{\rm{eff}}_{z,v+e}&=S^{-1/2}(H_{z}-H^{z,\sigma}_{SOC}H_{\sigma}^{-1}(H_{SOC}^{z,\sigma})^\dagger)S^{-1/2},\nonumber\\
&=S^{-1/2}H_z S^{-1/2}\nonumber\\
&-S^{-1/2}H^{z,\sigma}_{SOC}H_{\sigma}^{-1}(H_{SOC}^{z,\sigma})^\dagger S^{-1/2}.
\end{align}
In the second line, we have split the Hamiltonian in two parts. The first term on the right-hand side (RHS) can neither lead to the opening of a gap, nor can it shift the position of the Dirac cones. This follows from the relation $\det{[S^{-1/2}H_zS^{-1/2}]}=\det{H_z}/\det{S}$. Hence, as for the first term, we may simply set $S=\mathbb{I}$. With respect to the second term, we use
\begin{align}
S^{-1/2}&=\mathbb{I}-\frac{1}{2}H^{z,\sigma}_{SOC}H_\sigma^{-1}(H^{z,\sigma}_{SOC})^\dagger+\ldots
\end{align}
This shows that if we use $\mathbb{I}$, we neglect contributions proportional to $\xi^4$ in the second term, with $\xi\in\{\xi_{sp1},\xi_{sp2},\xi_{p1},\xi_{p2}\}$.  This is allowed, since all $\xi$'s are very small compared to the other hopping parameters. As a result, we approximate the effective Hamiltonian by
\begin{align}\label{eq18}
H^{\rm{eff}}_{z,v+e}&=H_{z}-H^{z,\sigma}_{SOC}H_{\sigma}^{-1}(H_{SOC}^{z,\sigma})^\dagger.
\end{align}
Since in this approximation the Hamiltonian is given in momentum-space, we need to perform an inverse Fourier transformation to obtain a real-space TB Hamiltonian. Using that the  hoppings which form the bonds are the largest energies in the system, we can simplify $H_\sigma^{-1}$. First of all, we use $sp$, $sp^2$, and $p$ hybrid orbitals,  shown in Figs.~\ref{fig:sigmalattices}(a), (b), and (c). In this model, we only take into account the on-site energies $\varepsilon_i$, on-site hoppings $V_5,\ldots,V_9$, as well as the NN hoppings $V_1,\ldots,V_4$ which form a bond, yielding
\begin{align}
H_\sigma&=H_{\sigma,\rm{onsite}}+H_{\sigma,\rm{NN}}.
\end{align}
\begin{figure}[t]
\centering
\includegraphics[width=.5\textwidth]{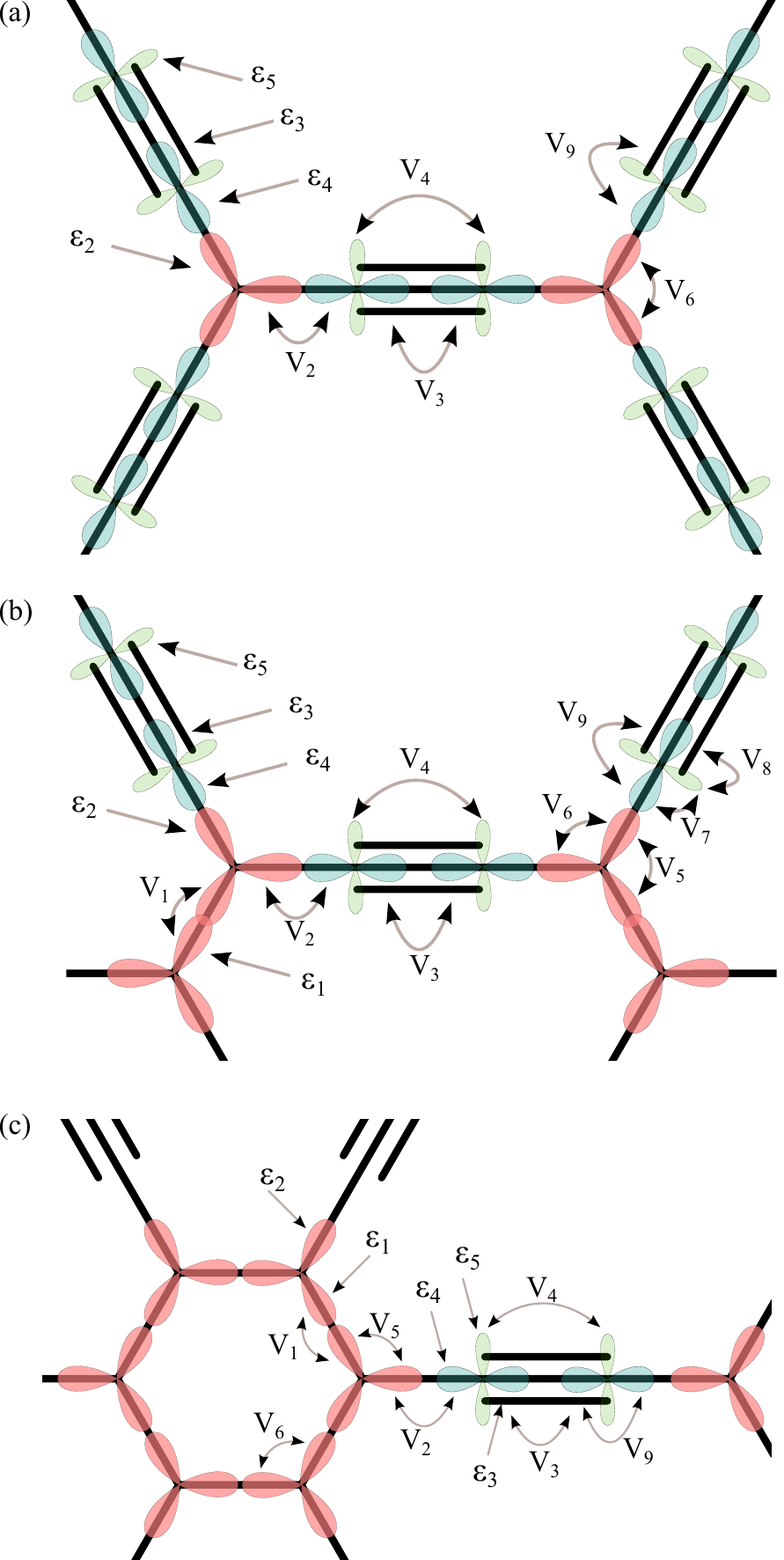}
\caption{\label{fig:sigmalattices} (Color online) Parameters used for the $\sigma$-TB models for $\alpha$, $\beta$, and $\gamma$-graphyne, shown in panels (a), (b), and  (c). The hopping parameters $V_1,\ldots,V_4$ correspond to NN hoppings, whereas $V_5,\ldots,V_9$ correspond to on-site hoppings, and $\varepsilon_1,\ldots,\varepsilon_5$ denote the on-site energies.  Other hopping parameters have been set to zero.}
\label{fig:sigmalattices}
\end{figure}
If the on-site energies and hopping parameters of the hybrid orbitals are smalller than the NN-hopping parameters of the bonds, we may approximate the inverse matrix by
\begin{align}\label{eqn18}
H_{\sigma}^{-1}\approx H_{\sigma,\rm{NN}}^{-1}-H_{\sigma,\rm{NN}}^{-1}H_{\sigma,\rm{onsite}}H_{\sigma,\rm{NN}}^{-1}.
\end{align}
This approximation is justified by the fact that the hybrid orbitals are mainly composed of $p$ orbitals that have very small on-site energies. In this simple TB model, we find that
\begin{align}
\left(H_{\sigma,\rm{NN}}^{-1}\right)_{ij}=\left\{
	\begin{array}{ll}
		1/\left(H_{\sigma,\rm{NN}}\right)^*_{ij}  & \mbox{if } (H_{\sigma,\rm{NN}})_{ij} \neq 0 \\
		0 & \mbox{if } (H_{\sigma,\rm{NN}})_{ij} = 0.
	\end{array}\right.
\end{align}
This expression can easily be transformed to real space. The effective SOC Hamiltonian is then given by the second term on the RHS of Eq.~(\ref{eq18}) combined with Eq.~(\ref{eqn18}),
\begin{align}\label{soceff}
H_{\rm{SOC, eff}}&=H^{z,\sigma}_{SOC}\left(H_{\sigma,\rm{NN}}^{-1}\right.\\
&\left.-H_{\sigma,\rm{NN}}^{-1}H_{\sigma,\rm{onsite}}H_{\sigma,\rm{NN}}^{-1}\right)(H_{SOC}^{z,\sigma})^\dagger.\nonumber
\end{align}
Using the Hamiltonians $H^{z,\sigma}_{SOC}$, $H_{\sigma,\rm{NN}}$, and $H_{\sigma,\rm{onsite}}$ for $\alpha$-, $\beta$-, and $\gamma$-graphyne, we obtain the SOC in real space
\begin{align}\label{soceff2}
H^\sigma_{\rm{SOC, eff}}&=H^\sigma_{\rm{R}}+H^\sigma_{\rm{I}}+H^\sigma_{\rm{rest}},
\end{align}
where
\begin{align}\label{eqn22}
H^\sigma_{\rm{R}}&=i\sum_{\langle i,j\rangle}\lambda^\sigma_{R,ij}p_{z,i}^\dagger\left({\bm\sigma}\times{\bf\hat{d}}_{ij}\right)\cdot{\bf\hat{z}}p_{z,j},\\\label{eqn23}
H^\sigma_{\rm{I}}&=i\sum_{\langle\langle i,j\rangle\rangle}\lambda^\sigma_{I,ij}v_{ij}p_{z,i}^\dagger \sigma_z p_{z,j},
\end{align}
and $H_{\rm{rest}}$ describes the next-nearest neighbor (NNN) corrections to the Rashba SOC (which can be neglected), together with some very small spin-independent and thus negligible NN and NNN hoppings and on-site energies. In both expressions, we have included bond-dependent coupling constants $\lambda^\sigma_{R,ij}$ and $\lambda^\sigma_{I,ij}$, ${\bf\hat{d}}_{ij}$ is the unit vector pointing from site $i$ to $j$, and $v_{ij}=+(-)$ if the hopping is (anti)-clockwise, and zero if it is along the acetylene bond. In total, we  need six different coupling parameters to completely describe SOC in graphyne: three for Rashba ($\lambda^\sigma_{R,i}$) and three for the intrinsic SOC ($\lambda^\sigma_{I,i}$), with $i=1,2,3$ corresponding to vertex-vertex,  vertex-edge, and edge-edge hoppings, respectively. In Table \ref{table1} we give the expressions for these coupling parameters in terms of the hopping parameters for the $\sigma$-orbitals, as obtained from Eqs.  (\ref{eqn22}) and (\ref{eqn23}).\\

It turns out that for the three types of graphyne that we consider, we obtain Hamiltonians of the same form as in graphene, simply with adjusted coupling parameters. However, in $\beta$-graphyne there is an additional contribution to the intrinsic SOC Hamiltonian. In $\alpha$- and $\gamma$-graphyne, each hopping along a straight path yields no contribution to the intrinsic SOC Hamiltonian, which is a consequence of the mirror symmetry through the acetylene bond in these systems. This symmetry is weakly broken in $\beta$-graphyne. As a result of this weak symmetry breaking, we obtain a very small additional contribution to the intrinsic SOC Hamiltonian,
\begin{align}\label{eqn24}
H_{\beta,4}&=i\lambda^\sigma_{I,4}\sum_{\langle\langle i,j\rangle\rangle}w_{ij}p^\dagger_{z,i}\sigma_z p_{z,j},
\end{align}
where $w_{ij}=+(-)$ if the hopping is along the acetylene bond going (anti-)clockwise with respect to the center of the unit-cell, and $\lambda^\sigma_{I,4}=\sqrt{2}\xi_{p1}\xi_{p2}V_7/(\sqrt{3}V_2 V_4)$. Since $V_7$ is non-zero only in $\beta$-graphyne, this  contribution is absent in $\alpha$- and $\gamma$-graphyne. Furthermore, we would like to point out that in the TB models that we used, we have neglected subdominant NN hoppings. However, the inclusion of these can lead to an NN intrinsic SOC term in the Hamiltonian. In $\beta$-graphyne, this is generated by the NN hopping from the $sp^2$ orbitals to the $p$ orbitals, whereas in $\gamma$-graphyne this is caused by the hopping between $sp^2$ orbitals which point in different directions. These contributions can easily be explained from the broken mirror symmetry through the acetylene bond in $\beta$-graphyne,  and through the $\sigma$-bond forming the hexagon in $\gamma$-graphyne. Since these symmetries are only weakly broken, their effect will be very small.\\

In Ref.\ [\onlinecite{MZhaoSciRep2013}], the effect of intrinsic SOC on the dispersion relation in graphyne was calculated from first principles. The predicted gap of $0.014$ meV in $\alpha$-graphyne is rather large as compared to graphene, and may be attributed  to the inhomogeneity in the charge distribution around the acetylene bond. By fitting the gap obtained within the TB model to the one derived from first-principles calculations, we find that for $\alpha$-graphyne $\lambda^\sigma_{I,3}\approx 0.041$meV. In $\alpha$-graphyne $V_6=\epsilon_s/3\approx 2.67$eV, and a rough estimate yields  $V_2\approx  5-10$eV. This leads to an approximate lower bound for $\xi_{p1}\approx  12.6$meV, whereas in graphene $\xi_p\approx 2.8$meV. In addition, around the vertices the charge is rather homogeneously distributed, and thus we expect that the coupling $\xi_{p2}$ is roughly of the same size as in graphene, i.e. $\xi_{p2}\approx2.8$meV.\\

Inspection of Table~\ref{table1} clearly shows that the parameters corresponding to the Rashba coupling are proportional to $\xi_{sp}\xi_{p}/V_i$, where $V_i$ is one of the NN hopping parameters. This is expected, since the Rashba terms arise from
\begin{align}\label{27}
H^{z,\sigma}_E H_{\sigma,\rm{NN}}^{-1}(H^{z,\sigma}_L)^\dagger+h.c.,
\end{align}
where the matrix $H^{z,\sigma}_E$ contributes a factor $\xi_{sp}$, $H_{\rm{NN}}$ yields a factor $V_i$, and $H^{z,\sigma}_L$ a factor $\xi_p$. Note that $V_4$ does not appear in Table~\ref{table1}; this can be understood from the matrix structure of Eq.~(\ref{27}). Terms from right to left in this expression correspond to {\it(i)} the on-site hopping from a $p_z$ orbital to a hybrid orbital, {\it (ii)} subsequent hopping between two NN hybrid orbitals, and {\it(iii)} the on-site hopping from an $sp$ or $sp^2$ hybrid orbital to a $p_z$ orbital due to the electric field. Since $V_4$ is responsible for the hopping between two NN $p$ orbitals, it does not contribute to this process. Notice also that the parameters $\varepsilon_1,...,\varepsilon_5$ and $V_5,...,V_9$ do not appear here because they correspond to on-site energies and hoppings, respectively.\\

The intrinsic SOC arises from
\begin{align}\label{eq1}
H^{z,\sigma}_L H_{\sigma,\rm{NN}}^{-1}H_{\sigma,\rm{onsite}}H_{\sigma,\rm{NN}}^{-1}(H^{z,\sigma}_L)^\dagger+h.c.;
\end{align}
hence, the coupling parameters are all of the form $\xi_p^2 A B^{-2}$, where $A\in\{V_5,\ldots,V_9,\varepsilon_1,\ldots,\varepsilon_5\}$ comes from $H_{\sigma,\rm{onsite}}$ and $B\in\{V_1,\ldots,V_4\}$ comes from $H_{\sigma,\rm{NN}}$. However, Table~\ref{table1} clearly shows that actually none of these parameters are proportional to an onsite energy $\varepsilon_i$. The reason for this can be understood as follows.  Reading Eq.~(\ref{eq1}) from right to left, we find that {\it (i)} the first matrix leads to the hopping from a $p_z$ orbital to one of the hybrid orbitals due to the SOC, {\it (ii)} then $H_{\sigma,\rm{NN}}^{-1}$ leads to the hopping to a NN hybrid orbital,{\it (iii,a)} the term $H_{\sigma,\rm{onsite}}$ can lead to the onsite hopping to another hybrid orbital; this contributes a factor $V_j$ with $j=5,\ldots,9$,{\it (iii,b)} or it can simply stay on the same orbital which would contribute a factor $\varepsilon_i$. Therefore, only this last scenario would yield a contribution proportional to $\varepsilon_i$. However, this scenario then subsequently leads to the hopping to the hybrid orbital  where one started; therefore there are no terms proportional to $\varepsilon_i$. The parameter $V_8$  is also absent in Table~\ref{table1}. This can be understood by analyzing  the hopping process proportional to $V_8$: starting from the $p_z$ orbital $a^1$ (see Appendix~\ref{c2}, Fig.~\ref{figa} therein,  and Fig.~\ref{fig:sigmalattices}),  we then find the following:
\begin{enumerate}
\item $(H^{z,\sigma}_{L})^\dagger$ contributes a factor proportional to $\sigma_y$, and leads to the hopping to state $a^1_1$.
\item $H_{\sigma,\rm{NN}}^{-1}$ contributes a factor $V_3^{-1}$, and leads to the hopping to state $b^1_1$.
\item $H_{\sigma,\rm{onsite}}$ contributes a factor $V_8$, and leads to the hopping to state $b^1_3$.
\item $H_{\sigma,\rm{NN}}^{-1}$ contributes a factor $V_3^{-1}$, and leads to the hopping to state $a^1_1$.
\item $(H^{z,\sigma}_{L})^\dagger$ contributes a factor proportional to $\sigma_y$, and leads to the hopping to the $p_z$-orbital $a^1$.
 \end{enumerate}
 Hence, the term proportional to $V_8$ does not lead to a NNN hopping process,  but gives rise to an onsite energy, which can be neglected. Concerning the absence of $V_9$ and $V_3$ in Table~\ref{table1} for the intrinsic SOC parameters, we would like to point out that these terms do actually lead to a NNN hopping in Eq.~(\ref{eq1}) that is, however, spin-independent. This can be seen from the fact that from right to left again, for this process the hybrid orbital to which the $p_z$ orbital hops points in the same direction as the hybrid orbital from which it hops to the NNN $p_z$ orbital. To illustrate this process, we consider the hopping from $A$ to $b^1$ via $V_9$ and $V_3$ (see Fig.~\ref{figa} and Fig.~\ref{fig:sigmalattices}):
\begin{enumerate}
\item $(H^{z,\sigma}_{L})^\dagger$ contributes a factor proportional to $\sigma_y$, and leads to the hopping to state $A_2$.
\item $H_{\sigma,\rm{NN}}^{-1}$ contributes a factor $1/V_2$, and leads to the hopping to state $a^1_2$.
\item $H_{\sigma,\rm{onsite}}$ contributes a factor $V_9$, and leads to the hopping to state $a^1_1$.
\item $H_{\sigma,\rm{NN}}^{-1}$ contributes a factor $1/V_3$, and leads to the hopping to state $b^1_1$.
\item $(H^{z,\sigma}_{L})^\dagger$ contributes a factor proportional to $\sigma_y$, and leads to the hopping to state $b^1$.
\end{enumerate}
Since the initial and final hoppings are both proportional to $\sigma_y$, we find that the combination is proportional to $\sigma_y\cdot\sigma_y=\mathbb{I}$; hence, it does not lead to a spin-dependent hopping. Note that the parameters $V_4$ and $V_7$ appear in the expression for $\lambda^\sigma_{I,4}$ in Eq.~(\ref{eqn24}).
\begin{table}[t]

\centering
\begin{tabular}{ l || c  }
   $\sigma$& $\alpha$, $\beta$, and $\gamma$ \\
  \hline\hline
  $\lambda_{R,1}^\sigma$ & $2\sqrt{2}\xi_{2sp}\xi_{2p}/(3V_1)$ \\
  $\lambda_{R,2}^\sigma$ & $(\sqrt{2}\xi_{sp1}\xi_{p2}+\xi_{sp2}\xi_{p1})/(\sqrt{6}V_2)$\\
  $\lambda_{R,3}^\sigma$ & $\xi_{1sp}\xi_{1p}/V_3$ \\
  $\lambda_{I,1}^\sigma$  & $V_6\xi_{p,2}^2/(\sqrt{3}V_1^2)$\\
  $\lambda_{I,2}^\sigma$  & $V_{5}\xi_{p,2}\xi_{p,1}/(2V_1 V_2)$\\
  $\lambda_{I,3}^\sigma$ & $\sqrt{3}V_6\xi_{p1}^2/(4V_2^2)$ \\
  $\lambda^\sigma_{I,4}$ & $\sqrt{2}\xi_{p1}\xi_{p2}V_7/(\sqrt{3}V_2 V_4)$\\
\end{tabular}
\caption{SOC parameters for the three types of graphyne arising from the $\sigma$-orbitals.}
\label{table1}
\end{table}
\subsection{Spin-orbit coupling generated by the $d$-orbitals.}
\begin{table}[t]
\centering
\begin{tabular}{ l || c  }
  d & $\alpha$, $\beta$, and $\gamma$ \\
  \hline\hline
  $\lambda_{I,1}^d$  & $\sqrt{3}V_{dp1}^2\xi_d/2\epsilon_d^2$\\
  $\lambda_{I,2}^d$  & $\sqrt{3}V_{dp1}V_{dp2}\xi_d/2\epsilon_d^2$\\
  $\lambda_{I,3}^d$ & $\sqrt{3}V_{dp2}^2\xi_d/2\epsilon_d^2$ \\
\end{tabular}
\caption{SOC parameters for the three types of graphyne arising from the $d$-orbitals.}
\label{table2}
\end{table}
The preceding discussion showed that the coupling parameters for the intrinsic SOC are of second order in $\xi_{p1}$ and (or) $\xi_{p2}$. In Ref.~[\onlinecite{fabian}], it was shown that the gap opening in graphene is actually due to intrinsic SOC hosted by the $d_{xz}$ and $d_{yz}$ orbitals. The reason is that the intrinsic SOC due to the $d$ orbitals is of first order in the intrinsic SOC parameter for the $d$-orbitals, $\xi_d$. To include the contributions stemming from the $d$-orbitals, we write the TB Hamiltonian with the hopping between the $p_z$, $d_{xz}$, and $d_{yz}$  orbitals included, and the intrinsic SOC among the $d$-orbitals
\begin{align}
H&=H_z +H_{zd}+H_{zd}^\dagger+H_d+H_{L}^d.
\end{align}
Here, $H_z$ describes the hopping between the $p_z$ orbitals, $H_{zd}$ describes the hopping from the $d$ orbitals to the $p_z$ orbitals, $H_d$ is the TB Hamiltonian describing the $d_{xz}$ and $d_{yz}$ orbitals, and $H_{L}^d$  describes the intrinsic SOC between the $d_{xz}$ and $d_{yz}$ orbitals. The latter term is given by
\begin{align}
H_L^d&=i\xi_{d1}\sum_{i}{}^{'}d_{yz,i}^\dagger\sigma_z d_{xz,i}+i\xi_{d2}\sum_{i}{}^{''}d_{yz,i}^\dagger\sigma_z d_{xz,i}+h.c.
\end{align}
The Hamiltonian $H_{zd}$ is given by
\begin{align}
H_{zd}&=\sum_{\langle i,j\rangle}V_{dp,ij}p^\dagger_{z,i}(\hat{{\bf d}}_{ij}^x d_{xz,j}+\hat{{\bf d}}_{ij}^yd_{yz,j}),
\end{align}
where $\hat{{\bf d}}_{ij}^\mu$ is the $\mu$-component of the vector $\hat{{\bf d}}_{ij}$, defined below Eq.~(\ref{eqn23}) and $V_{dp,ij}=V_{dp,k}$ with $k=1,2,3$ depending whether $\hat{{\bf d}}_{ij}$ points from vertex to vertex, vertex to edge or edge to edge, respectively. Again, we use the approximation scheme from Appendix~\ref{eff} that leads to the following effective Hamiltonian:
\begin{align}
H^{\rm{eff}}_{z,v+e}&=S^{-1/2}[H_{z}-H_{zd}\left(H_{L}^d+H_d\right)^{-1}H_{zd}^\dagger]S^{-1/2}.
\end{align}
Following the  same reasoning as before, we simply set $S=\mathbb{I}$. Since $\xi_{d1}$ and $\xi_{d2}$ are very small, we may approximate $\left(H_{L}^d+H_d\right)^{-1}\approx H_d^{-1}-H_{d}^{-1}H_{L}^d H_{d}^{-1}$. Notice that this explains why the intrinsic SOC generated by the $d$-orbitals is first order in $\xi_d$. As a result, the effective intrinsic SOC Hamiltonian due to the $d$-orbitals is given by
\begin{align}\label{isod}
H_{\rm SOC,eff}=H_{zd}H_{d}^{-1}H_{L}^d H_{d}^{-1}H_{zd}^\dagger.
\end{align}
Because of the large on-site energies of the $d$ orbitals $\varepsilon_d$, we may approximate $H_{d}^{-1}\approx \varepsilon_d^{-1}\mathbb{I}$. Therefore, Eq.~(\ref{isod}) reduces to
\begin{align}
H_{\rm SOC,eff}=H_{zd}H_{L}^dH^\dagger_{zd}/\varepsilon_d^2.
\end{align}
Using the Hamiltonians $H_{L}^d$ and $H_{zd}$, we finally obtain the intrinsic SOC Hamiltonian due to the $d$-orbitals,
\begin{align}
H^d_{\rm{I}}&=i\sum_{\langle\langle i,j\rangle\rangle}\lambda^d_{I,ij}v_{ij}p_{z,i}^\dagger \sigma_z p_{z,j},
\end{align}
where $v_{ij}=(+)-$ if the hopping is (anti)-clockwise. As for the $\sigma$-orbitals, we will need three different coupling parameters ($\lambda^d_{I,i}$) to completely describe intrinsic SOC; the labeling used is the same as for the $\sigma$-orbitals. The coupling parameters are given in Table~\ref{table2}. Note that the expressions for the SOC parameters due to the $d$-orbitals and due to the $\sigma$-orbitals have the same sign. Therefore, the intrinsic SOC is governed by the parameters $\lambda_{I,j}=\lambda^\sigma_{I,j}+\lambda^d_{I,j}$. As a result, the inclusion of $d$-orbitals increases the effect of SOC in graphynes.\\

In Ref.~[\onlinecite{fabian}], it was shown that the effect of $d$-orbitals on the Rashba SOC is negligible. We expect that this remains true in graphynes, and for this reason we have not considered the effect of $d$-orbitals on the Rashba SOC. 
\subsection{Spin-orbit Hamiltonians in the two-site and six-site model}
\begin{table*}[t]
\centering
\begin{tabular}{ l || c  }
  SOC & $\alpha$, $\beta$, and $\gamma$\\
  \hline\hline
  $\tilde{\lambda}_{\rm I}^\alpha$ & $\lambda_{I,3}t_2^2/(t_3^2+3t_2^2)$\\
  $\lambda_{\rm ext,I}^\beta$  & $-\lambda_{I,2}t_2 t_3 /(t_3^2+2t_2^2)$\\
  $\lambda_{\rm ext,I}^\gamma$ & $-\lambda_{I,2}t_3t_2/(t_2^2+t_3^2)$\\
  $\lambda_{\rm int,I}^\beta$ & $\lambda_{I,3}t_2^2/(t_3^2+2t_2^2)$\\
  $\lambda_{\rm int,I}^\gamma$ & $\lambda_{I,1}t_3^2/(t_2^2+t_3^2)$\\
  $\tilde{\lambda}_{\rm R}^\alpha$ & $2\lambda_{R,2}t_2 t_3/(t_3^2+3t_2^2)+\lambda_{R,3}t_2^2/(t_3^2+3t_2^2)$\\
  $\lambda_{\rm ext,R}^\beta$ & $\lambda_{R,1}t_3^2/(t_3^2+2t_2^2)$\\
  $\lambda_{\rm ext,R}^\gamma$ & $2 \lambda_{R,2}t_2t_3/(t_3^2+t_2^2)+\lambda_{R,2}t_2^2/(t_3^2+t_2^2)$\\
  $\lambda_{\rm int,R}^\beta$ & $-2\lambda_{R,2}t_2t_3/(t_3^2+2t_2^2)-\lambda_{R,3}t_2^2/(t_3^2+2t_2^2)$\\
  $\lambda_{\rm int,R}^\gamma$ & $\lambda_{R,1}t_3^2/(t_2^2+t_3^2)$\\
\end{tabular}
\caption{SOC parameters for the effective TB models.}
\label{table8}
\end{table*}
The SOC Hamiltonians that we obtained are all given in terms of the full TB Hamiltonian. The next step is to integrate out the high-energy orbitals to obtain SOC Hamiltonians that can be used in the effective models introduced in Sec.~II. In general, the effective Hamiltonian reads as
\begin{align}\label{effsoc}
H^{\rm{eff}}_{z,v}&=S^{-1/2}\left(H_{vv}-H_{ve}H_{ee}^{-1}H_{ve}^\dagger\right)S^{-1/2},
\end{align}
where $H_{vv}$ ($H_{ee}$) describe the orbitals at the vertices (edges), and  $H_{ve}$ mixes them. In order to incorporate SOC in this description, we write each matrix as the sum of a spin-independent part, denoted by a tilde, and a part describing the SOC, denoted by the subscript SOC. Because the SOC parameters are very small compared to the other hopping energies, we may expand Eq.~(\ref{effsoc}) up to first order in them. One then readily obtains
\begin{align}
S^{-1/2}&=(\tilde{S}+S_{SOC})^{-1/2},\nonumber\\
&\approx \tilde{S}^{-1/2}-\frac{1}{2}\tilde{S}^{-3/2}S_{SOC},
\end{align}
where
\begin{align}
S_{SOC}&=H_{ve,SOC}\tilde{H}_{ee}^{-2}\tilde{H}_{ve}^\dagger\nonumber\\
&-\tilde{H}_{ve}\tilde{H}_{ee}^{-2}H_{ee,SOC}\tilde{H}^{-1}_{ee}\tilde{H}_{ve}^\dagger+h.c.
\end{align}
However, if we simply set $S=\tilde{S}$ we do not miss any gap openings or shifts in the positions of the Dirac cones. Hence, we approximate Eq.~(\ref{effsoc}) by
\begin{align}
H^{\rm{eff}}_{z,v}\approx\tilde{S}^{-1/2}\left(H_{vv}-H_{ve}H_{ee}^{-1}H_{ve}^\dagger\right)\tilde{S}^{-1/2}.
\end{align}
As a result, we find
\begin{align}
H^{\rm{eff}}_{z,v}&\approx\tilde{H}^{\rm{eff}}_{z,v}+H_{SOC},
\end{align}
where $H_{SOC}=H_{1,SOC}+H_{2,SOC}+H_{3,SOC}$, with
\begin{align}
H_{1,SOC}&=\tilde{S}^{-1/2}H_{vv,SOC}\tilde{S}^{-1/2},\\
H_{2,SOC}&=-\tilde{S}^{-1/2}H_{ve,SOC}\tilde{H}_{ee}\tilde{H}_{ve}^\dagger\tilde{S}^{-1/2}+h.c.,\\
H_{3,SOC}&=\tilde{S}^{-1/2}\tilde{H}_{ve}\tilde{H}_{ee}^{-1}H_{ee,SOC}\tilde{H}_{ee}^{-1}\tilde{H}_{ve}^\dagger.
\end{align}
By performing these calculations for $\alpha$-graphyne, we obtain
\begin{align}\label{aiso}
H_{R,\alpha}&=i\lambda_{R}\sum_{\langle i,j\rangle}p_{z,i}^\dagger\left({\bm \sigma}\times \hat{{\bf d}}_{ij}\right)\cdot\hat{{\bf z}} p_{z,j},\\
H_{I,\alpha}&=i\lambda_{I}\sum_{\langle\langle i,j\rangle\rangle}v_{ij}p_{z,i}^\dagger\sigma_z p_{z,j},\label{arashba}
\end{align}
which are the standard SOC Hamiltonians, as used for graphene. The results for $\beta$- and $\gamma$-graphyne are slightly different than for graphene, since now we have to distinguish between the inter and intra-unit cell SOC. We refer to the inter-unit cell SOC as external SOC, and to the intra unit cell SOC as internal SOC. The form of the SOC Hamiltonians is however unchanged as compared to graphene,
\begin{align}\label{intext}
H_{R,\beta/\gamma}&=i \lambda_{\rm int,R}\dot{\sum_{\langle i,j\rangle}}p_{z,i}^\dagger \left({\bm \sigma}\times \hat{{\bf d}}_{ij}\right)\cdot\hat{{\bf z}} p_{z,j}\\
&+i \lambda_{\rm ext,R}\ddot{\sum_{\langle i,j\rangle}}p_{z,i}^\dagger \left({\bm \sigma}\times \hat{{\bf d}}_{ij}\right)\cdot\hat{{\bf z}} p_{z,j},\nonumber\\
H_{I,\beta/\gamma}&=i \lambda_{\rm int,I}\dot{\sum_{\langle\langle i,j\rangle\rangle}}v_{ij}p_{z,i}^\dagger\sigma_z p_{z,j}\label{intext2}\\
&+i \lambda_{\rm ext,I}\ddot{\sum_{\langle\langle i,j\rangle\rangle}}v_{ij}p_{z,i}^\dagger\sigma_z p_{z,j}.
\end{align}
Here, the single (double) dot indicates that the sum is taken over sites within the same (belonging to different) unit cells. In Table~\ref{table8} we have listed the effective SOC hopping parameters.

\section{Internal and external spin-orbit couplings}
As shown in Sec.~II, both $\beta$- and $\gamma$-graphyne can be described in terms of the same six-site model, and Eqs.~(\ref{effb}) and (\ref{effc}) can be rewritten in a short-hand notation as
\begin{align}\label{efff}
H_0&=t_{\rm int}\dot{\sum_{\langle i,j\rangle}}p_{z,i}^\dagger p_{z,j}+t_{\rm ext}\ddot{\sum_{\langle i,j\rangle}}p_{z,i}^\dagger p_{z,j}.
\end{align}
The band structure  obtained from this Hamiltonian exhibits six Dirac cones at the line connecting the $\Gamma$ and $M$ points if the condition $-2<t_{\rm ext}/t_{\rm int}<-1$ is satisfied. This is realized for $\beta$- but not for $\gamma$-graphyne [see Figs.~\ref{fig:dispersion2}(b) and \ref{fig:dispersion2}(c)]. We show in the following that the intrinsic SOC can open a non-trivial gap in $\beta$-graphyne, whereas the Rashba SOC can be used to open or close a trivial gap in $\beta$- and $\gamma$-graphyne.  In Eqs.~(\ref{intext}) and (\ref{intext2}), we have made a distinction between external and internal SOC. To obtain a better understanding of the effect of both terms, we discuss them separately in the following.
\subsection{Internal Rashba spin-orbit coupling}
\begin{figure}[t]
\centering
\includegraphics[width=0.5\textwidth]{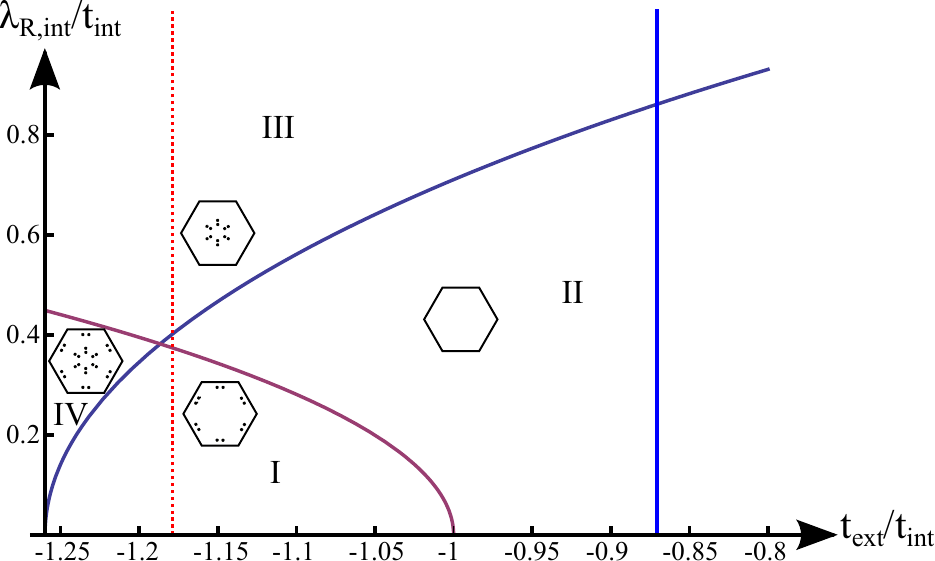}
\caption{(Color online) Phase diagram for the effective Hamiltonian~(\ref{efff}) with internal Rashba SOC. In region I the  system exhibits 12 Dirac cones, region II corresponds to the gapped phase, region III corresponds to the system where only pairs of Dirac cones along the line $\Gamma$-$M$ are present,  and region IV corresponds to a system where there are six pairs of cones along the line $\Gamma$-$M$ and six pairs on lines perpendicular to $\Gamma$-$M$. }
\label{fig:2.8}
\end{figure}

The internal Rashba SOC leads to very interesting phases characterized by the presence of Dirac cones at different points in the BZ. First, we discuss the regime that applies to $\beta$-graphyne, after which we consider the regime that describes $\gamma$-graphyne. At the end we also comment on the other regimes. Part of the phase diagram is shown in Fig.~\ref{fig:2.8}. The red dashed line therein corresponds to $\beta$-graphyne, whereas the blue solid line corresponds to $\gamma$-graphyne.\\

For  the regime $-1.19<t_{ext}/t_{int}<-1$ that applies to $\beta$-graphyne, the system goes through three different phases. Initially, the internal Rashba SOC splits each Dirac cone into a pair of Dirac cones located at a line perpendicular to the line connecting the $\Gamma$-$M$ points (see region I in Fig.~\ref{fig:2.8}). Upon increasing the coupling, these pairs move towards the boundary of the BZ, where they eventually annihilate with another pair at the line connecting the $K$ and $K'$ points. As a result, the system becomes gapped (see region II in Fig.~\ref{fig:2.8}). If the internal Rashba SOC is even further increased, the system undergoes another phase transition, with six new pairs of Dirac cones emerging along the lines connecting the $\Gamma$ and $M$ points (see region III in Fig.~\ref{fig:2.8}).\\

Next, we discuss the regime $-1<t_{ext}/t_{int}<-0.8$ that describes $\gamma$-graphyne. The system is initially gapped (see also region II in Fig.~\ref{fig:2.8}). However, when the internal Rashba SOC is sufficiently large, six pairs of Dirac cones appear along the lines connecting the $\Gamma$-$M$ points (see region III in Fig.~\ref{fig:2.8}).\\

\begin{figure}[b]
\centering
\subfigure[]{
\includegraphics[height=0.12\textwidth]{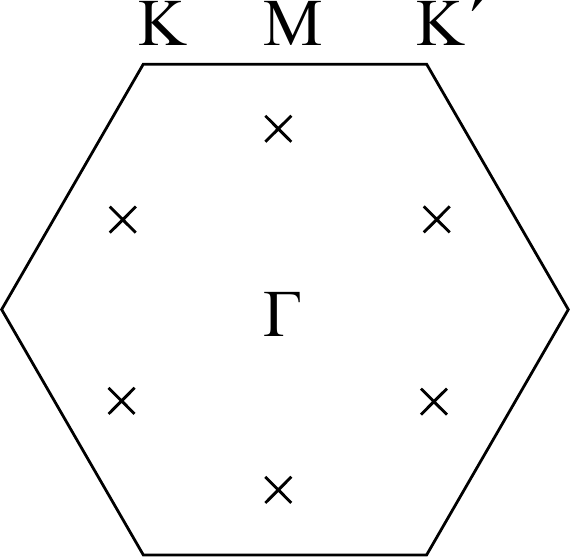}}
\subfigure[]{
\includegraphics[height=0.12\textwidth]{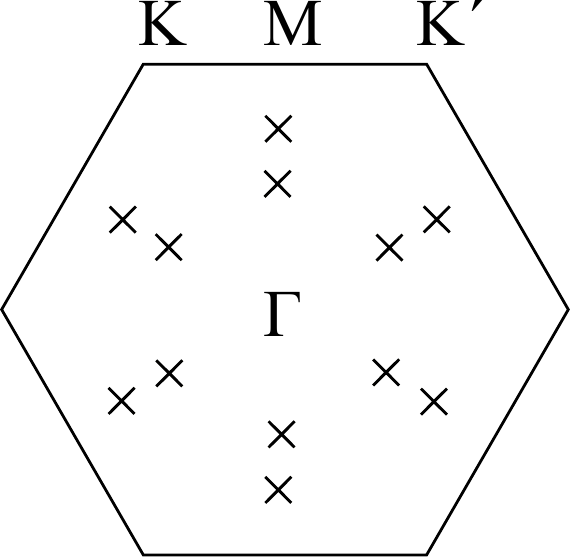}}
\subfigure[]{
\includegraphics[height=0.12\textwidth]{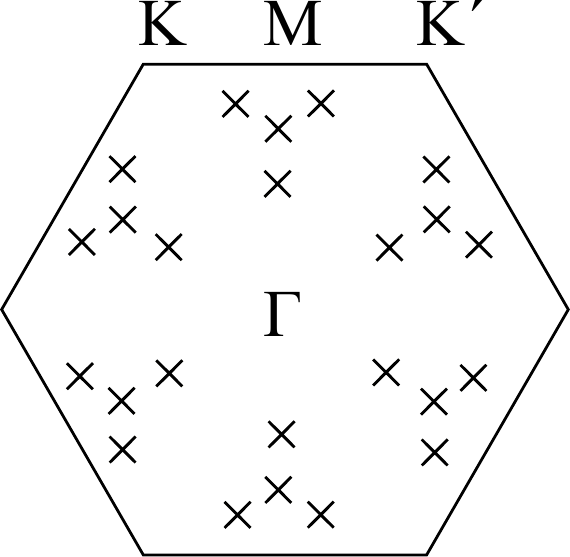}}
\caption{\label{fig:curious} Sketches of the phases corresponding to the regime $t_{\rm ext}/t_{\rm int}<-2^{1/3}$. Each cross denotes a Dirac cone and we have taken $t_{\rm ext}/t_{\rm int}=-1.3$. (a)  Brillouin zone for $\lambda_{\rm int,R}=\lambda_{\rm ext,R}=0$, (b)  Brillouin zone for $\lambda_{\rm int,R}/t_{\rm int}=0.1$ or $\lambda_{\rm ext,R}/t_{\rm int}=0.1$, and (c)  Brillouin zone for $\lambda_{\rm int,R}/t_{\rm int}=.4$ or $\lambda_{\rm ext,R}/t_{\rm int}=0.8$.}
\end{figure}

Having studied the regimes that apply to $\beta$- and $\gamma$-graphyne, we now consider the regime for which $-1.26<t_{ext}/t_{int}<-1.19$. As for $\beta$-graphyne, initially the internal Rashba SOC splits each Dirac cone into a pair of Dirac cones located along a line perpendicular to $\Gamma$-$M$ (see region I in Fig.~\ref{fig:2.8}). When the internal Rashba SOC is even further increased, six additional pairs of Dirac cones emerge along the line connecting $\Gamma$-$M$ (see region IV in Fig.~\ref{fig:2.8}). Eventually, when the coupling is even further increased, the six pairs of Dirac cones located at lines perpendicular to the $\Gamma$-$M$ points annihilate at the boundary of the BZ (see region III in Fig.~\ref{fig:2.8}). Finally, when $-2<t_{ext}/t_{int}<-1.26$, another curious phenomenon occurs (not shown in Fig.~\ref{fig:2.8}). First, each Dirac cone splits into a pair along the line connecting the $\Gamma$ and $M$ points [see Figs.~\ref{fig:curious}(a) and \ref{fig:curious}(b)]. When the coupling is even further increased, each of the cones closest to the $M$ points splits into  three cones [see Fig.~\ref{fig:curious}(c) for a sketch of the situation].

\subsection{External Rashba SOC}
 \begin{figure}[t]
\centering
\includegraphics[width=0.5\textwidth]{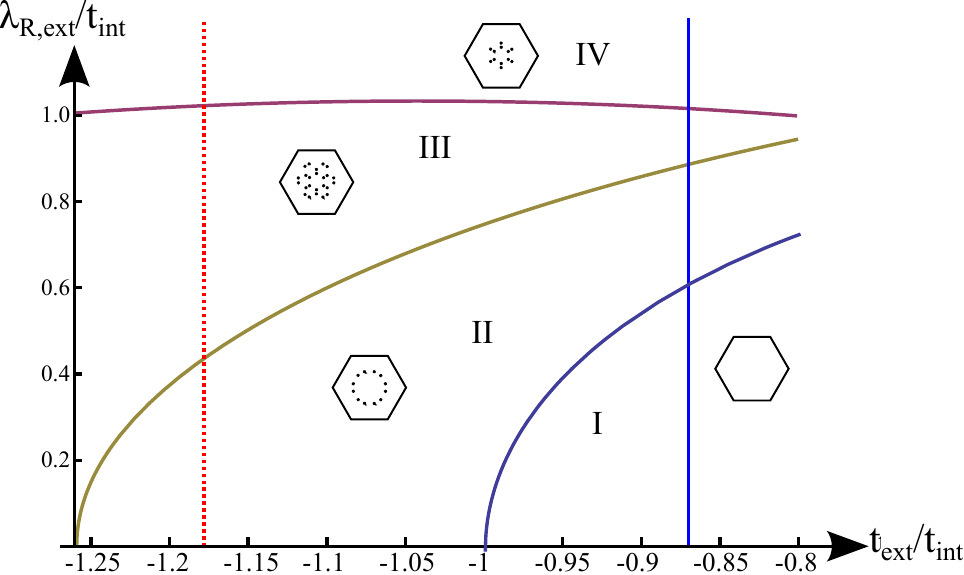}
\caption{(Color online) Phase diagram for the effective Hamiltonian~(\ref{efff}) with external Rashba SOC. Region I corresponds to a gapped system, region II exhibits six pairs of Dirac cones along a line perpendicular to the line connecting $\Gamma$ and $M$, region III exhibits six Dirac cones along a line perpendicular to the line connecting $\Gamma$ and $M$, and six pairs of Dirac cones along the lines connecting $\Gamma$ and $M$, and region IV shows the system with only six pairs of Dirac cones along the line connecting $\Gamma$ and $M$.}
\label{fig:2.9}
\end{figure}
As for the internal Rashba SOC, we first discuss the regimes that apply to $\beta$- and $\gamma$-graphyne. At the end, we shortly comment on the other regimes. The relevant phase diagram is shown in Fig.~\ref{fig:2.9}, where again the red dashed line corresponds to $\beta$-graphyne and the blue solid line corresponds to $\gamma$-graphyne.\\

First, we consider the regime $-1.26<t_{ext}/t_{int}<-1$ that applies to $\beta$-graphyne. Initially, the system exhibits six Dirac cones along the lines  connecting the $\Gamma$ and $M$ points. As the external Rashba SOC is switched on, each of these Dirac cones splits into a pair of Dirac cones located along lines perpendicular to the line connecting the $\Gamma$ and $M$ points (see region II in Fig.~\ref{fig:2.9}). At an intermediate value of the external Rashba SOC parameter, six additional pairs of Dirac cones emerge, located along the lines connecting the $\Gamma$ and $M$ points. When the coupling is even further increased, the pairs perpendicular to the lines connecting the $\Gamma$ and $M$ points eventually merge and vanish  along the lines connecting the $K$ ($K'$) and $\Gamma$ points (see region IV in Fig.~\ref{fig:2.9}).\\

We now move on to discuss the case $-1<t_{ext}/t_{int}<-0.8$,  the relevant regime for $\gamma$-graphyne. For $-1<t_{ext}/t_{int}$ the system is gapped, and remains to be such for small values of the external Rashba SOC (see region I in Fig.~\ref{fig:2.9}). However, at an intermediate value of the coupling, six pairs of Dirac cones emerge along the line connecting the $K$ and $K'$ points (see region II in Fig.~\ref{fig:2.9}). For increasing values of the coupling parameter $\gamma$-graphyne undergoes the same phase transitions as $\beta$-graphyne. Hence, subsequently it will enter regions III and IV (see Fig.~\ref{fig:2.9}). Finally,  in the regime $-2<t_{ext}/t_{int}<-1.26$ (not shown in Fig.~\ref{fig:2.9}) the external Rashba SOC acts in the same way as the internal Rashba SOC does (see also Fig.~\ref{fig:curious}).
\subsection{Interplay between internal and external Rashba spin-orbit coupling}
\begin{figure}[b]
\centering
\includegraphics[width=0.48\textwidth]{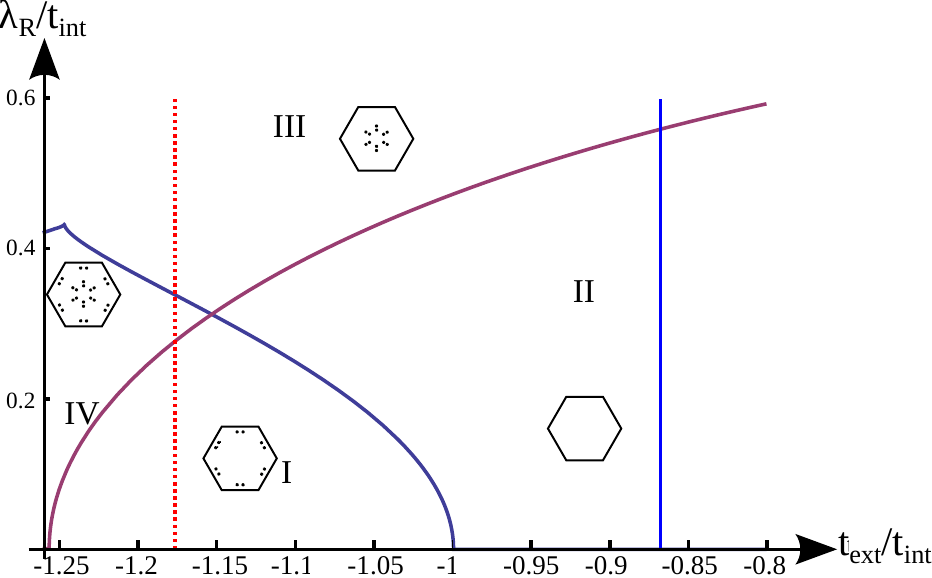}
\caption{(Color online) Phase diagram for the effective Hamiltonian~(\ref{efff}) with internal and external Rashba SOC, where $\lambda_R=\lambda_{\rm ext,R}=-\lambda_{\rm int,R}$. In region I the  system exhibits 12 Dirac cones, region II corresponds to the gapped phase, region III corresponds to the system where only pairs of Dirac cones along the line $\Gamma$-$M$ are present,  and region IV corresponds to a system where there are six pairs of cones along the line $\Gamma$-$M$ and six pairs on lines perpendicular to $\Gamma$-$M$. }
\label{fig:fase3}
\end{figure}

Since in real graphynes both the internal and external Rashba SOC are present simultaneously, we now consider this case. Inspection of Table~\ref{table8} shows that the internal Rashba SOC parameter $\lambda_{\rm ext,R}$ has an opposite sign compared to $\lambda_{\rm int,R}$. As a result, we study the case where the two parameters have opposite sign, and for simplicity we set their magnitudes to be equal. The relevant phase diagram is shown in Fig.~\ref{fig:fase3}. We observe that this phase diagram with  both couplings looks almost identical to the one for internal Rashba SOC only (compare Figs.~\ref{fig:2.8} and~\ref{fig:fase3}). However, there are some distinct features in the phase diagram containing both couplings. First of all, in Fig.~\ref{fig:2.8} the red (dashed) line, corresponding to $\beta$-graphyne crosses region II, whereas in Fig.~\ref{fig:fase3} this line crosses region IV. Hence, in this particular setup, the Rashba SOC does not open a gap in $\beta$-graphyne. Moreover, the line separating phases III and IV exhibits a cusp around $t_{\rm ext}/t_{\rm int}=-1.25$. It turns out that to the right of this cusp the Dirac cones merge along the lines connecting $K$ and $K'$ points, whereas to the left of this cusp the Dirac cones annihilate along the lines connecting $K$ and $\Gamma$ points. It should be noted that this latter behavior was also observed in the presence of external Rashba SOC (see Fig.~\ref{fig:2.9}). Most importantly, this shows that upon including both couplings the phase diagram interpolates between the cases when only one of the couplings is present (see Figs.~\ref{fig:2.8} and~\ref{fig:2.9}).
 \begin{figure}[t]
\centering
\subfigure[]{
\includegraphics[height=0.1415\textwidth]{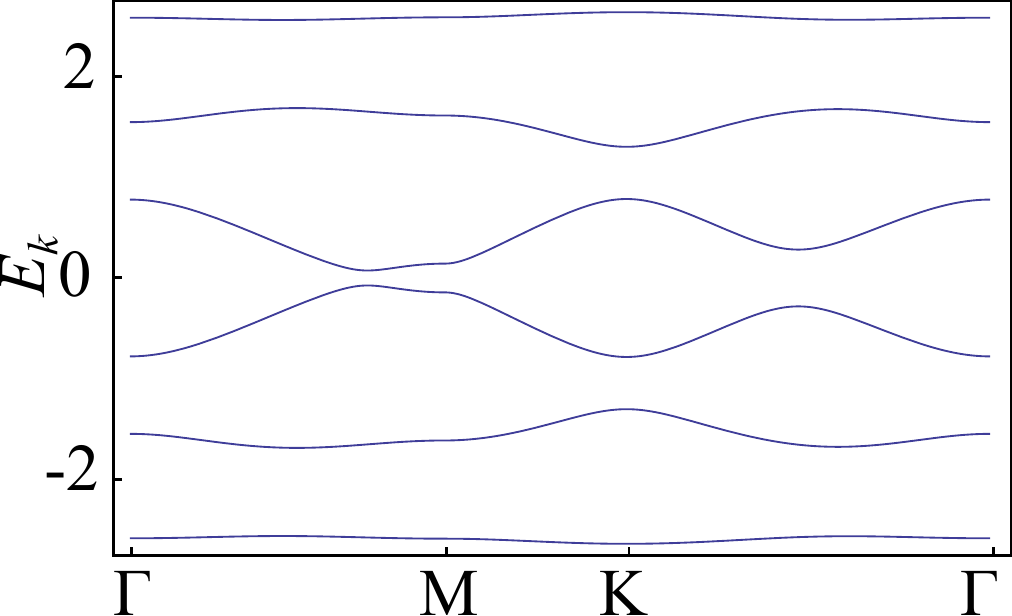}}
\subfigure[]{
\includegraphics[height=0.1415\textwidth]{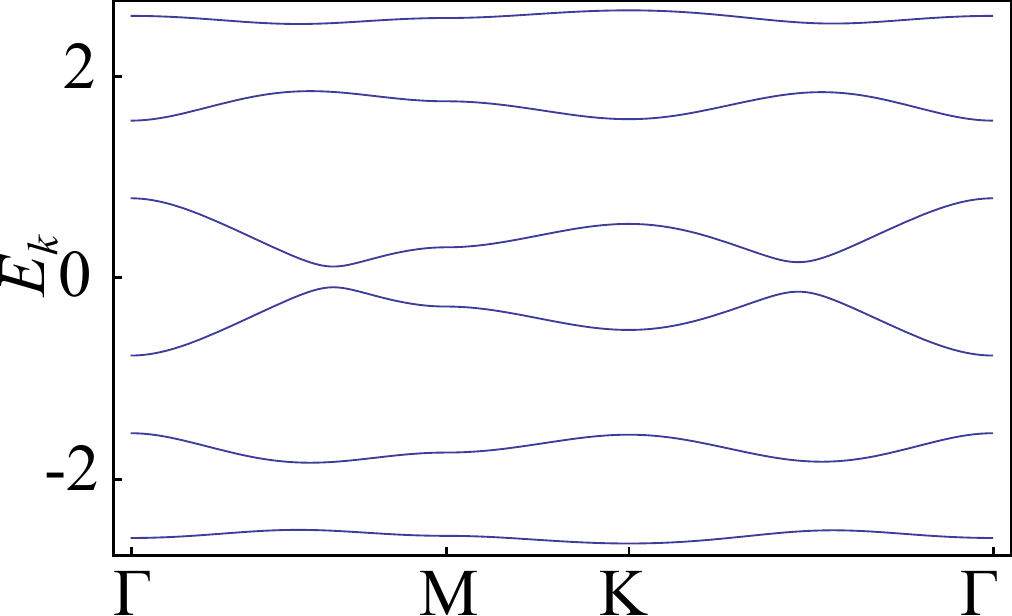}}
\subfigure[]{
\includegraphics[height=0.1415\textwidth]{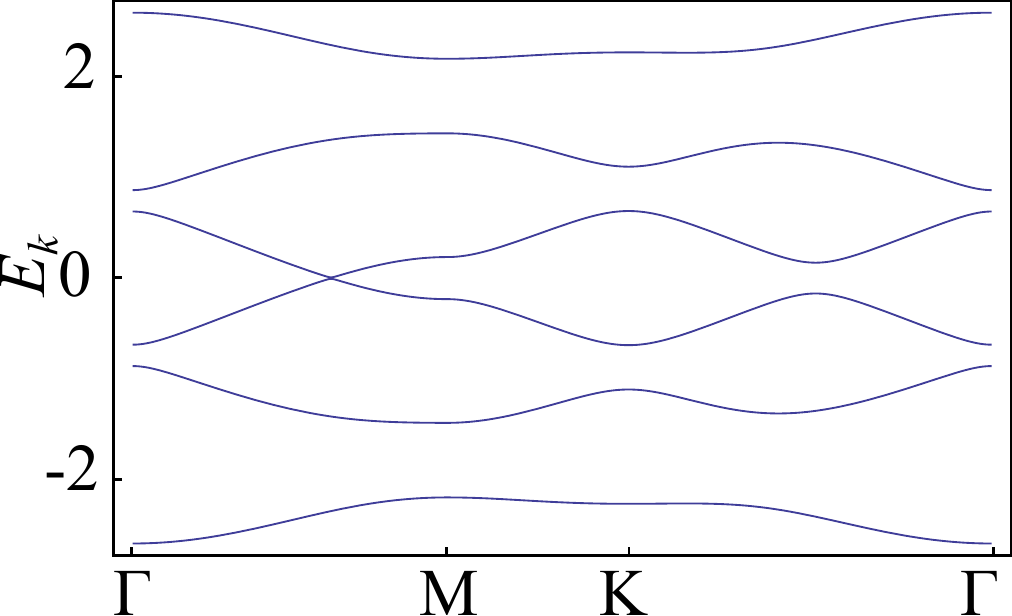}}
\subfigure[]{
\includegraphics[height=0.1415\textwidth]{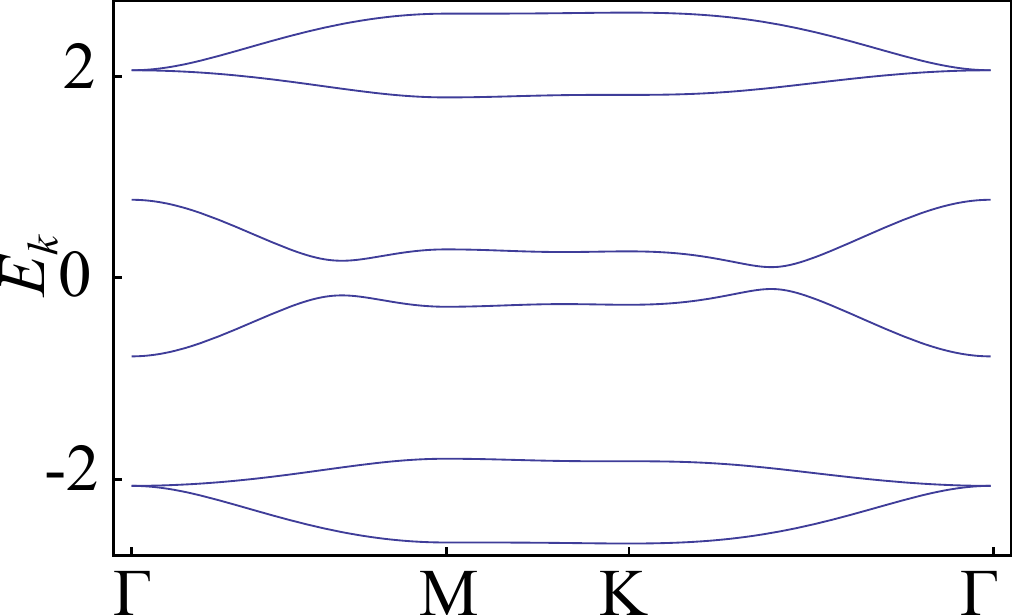}}
\caption{\label{fig:isoplusmin}(Color online) Band structure for $\beta$-graphyne for different values of $\lambda_{\rm int,I}$ and $\lambda_{\rm ext,I}$, along the line $k_x=0$. (a) $\lambda_{\rm ext,I}=0.15$eV and $\lambda_{\rm int,I}=0$eV. (b) $\lambda_{\rm ext,I}=0$eV and $\lambda_{\rm int,I}=0.3$eV. (c)  $\lambda_{\rm ext,I}=0.15$eV and $\lambda_{\rm int,I}=0.3$eV. (d)  $\lambda_{\rm ext,I}=0.15$eV and $\lambda_{\rm int,I}=-0.3$eV.}
\end{figure}
\subsection{Intrinsic spin-orbit coupling}
Whereas the internal and external Rashba SOC lead to a qualitative difference in the band structure, such a difference is absent when studying the intrinsic SOC. It is found that if the system exhibits Dirac cones, i.e. $-2<t_{\rm ext}/t_{\rm int}<-1$, both the internal and external intrinsic SOC open the gap located between the $\Gamma$ and $M$ points [see Figs.~\ref{fig:isoplusmin}(a) and~\ref{fig:isoplusmin}(b)], that turns out to be topologically nontrivial [\onlinecite{MMJ}]. However, if we combine both internal and external intrinsic SOC, both the magnitudes and the signs of the couplings $\lambda_{\rm int,I}$ and $\lambda_{\rm ext,I}$ play an important role. It turns out that if both couplings have the same sign, the two contributions tend to cancel each other, see Fig.~\ref{fig:isoplusmin}(c). On the other hand, if both couplings have opposite sign, the two contributions enhance the gap located at the line connecting the $\Gamma$ and $M$ points, see Fig.~\ref{fig:isoplusmin}(d).

Although at first sight  it might be surprising that for opposite sign of the coupling parameters $\lambda_{\rm int,I}$ and $\lambda_{\rm ext,I}$ the gap is enhanced, this results from the  fact that we have eliminated the $p_z$-orbitals located at the edges. For $\beta$-graphyne with orbitals at the edges included, the intrinsic SOC is governed by the coupling parameters $\lambda_{I,2}$ and $\lambda_{I,3}$ that have both the same sign, see Table~\ref{table1}. Then, if we eliminate the $p_z$-orbitals located at the edges, the intrinsic SOC is governed by the coupling parameters $\lambda_{\rm int,I}$ and $\lambda_{\rm ext,I}$. However, inspection of Table~\ref{table8} tells us that the coupling parameter describing the internal intrinsic SOC has an overall minus sign, whereas the external intrinsic SOC has not. Hence, a sign difference in the value for the parameters $\lambda_{\rm int,I}$ and $\lambda_{\rm ext,I}$ corresponds to the situation that in the full model, containing both orbitals at edges and vertices, the parameters $\lambda_{I,2}$ and $\lambda_{I,3}$ have the same sign.

\begin{figure}[t]
\centering
\includegraphics[width=0.4\textwidth]{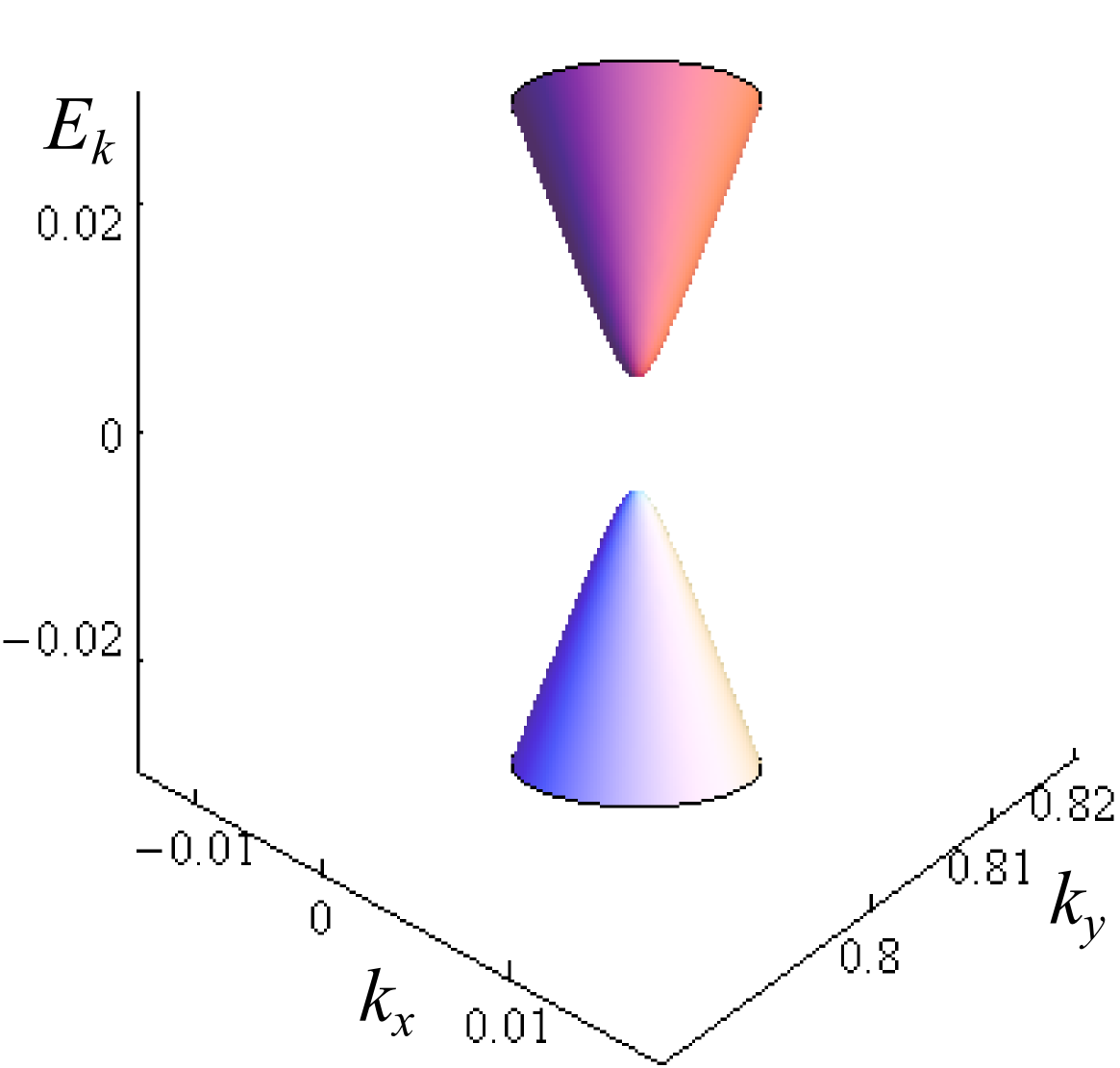}
\caption{\label{fig:alphametiso0,1}(Color online) Topological band gap opening at the $K$-point in $\alpha$-graphyne due to the intrinsic SOC, for $\lambda_{I}\approx 0.05$eV.}
\end{figure}
\section{Discussion and Conclusions}
\begin{figure}[b]
\includegraphics[width=0.4\textwidth]{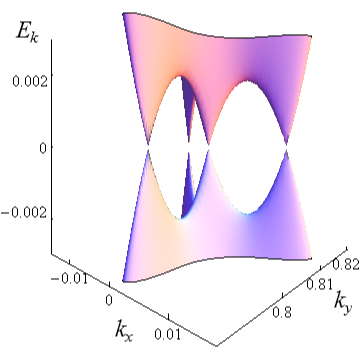}
\caption{\label{fig:alpharashba}(Color online) Band structure for $\alpha$-graphyne including the Rashba SOC for $\lambda_{R}\approx 0.5$eV, zoomed in on the $K$ point.}
\end{figure}
\begin{figure*}[t]
\centering
\includegraphics[width=\textwidth]{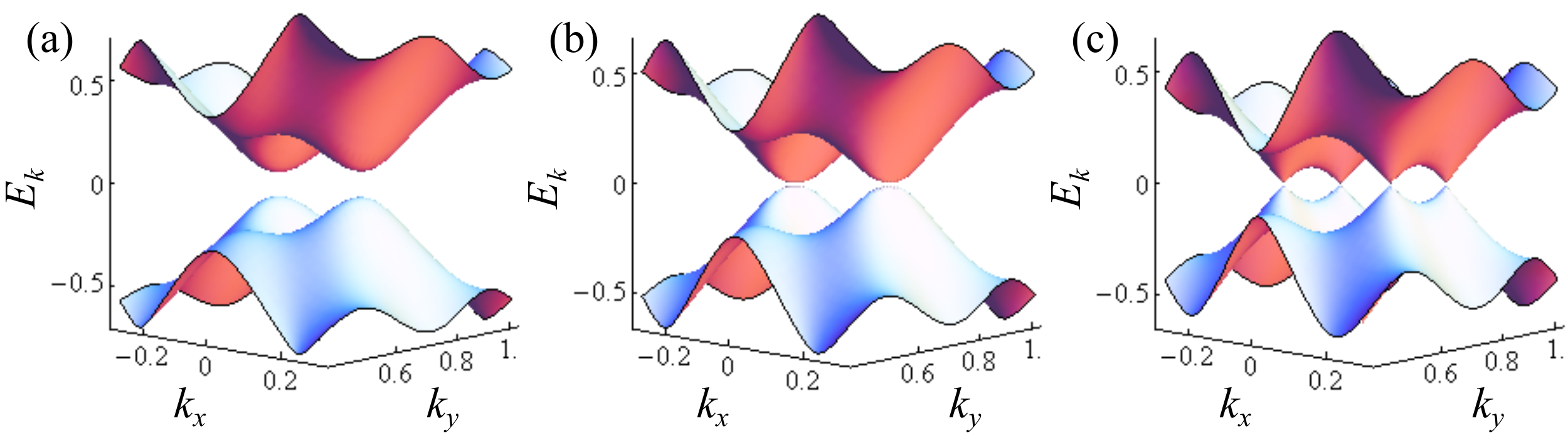}
\caption{\label{fig:merging} (Color online) Band structure of $\gamma-$graphyne in the presence of external Rashba SOC, zoomed in on the M-point in the BZ.  (a) Gapped system, for $\lambda_{\rm ext,R}\approx 0.9$eV, (b) transition from gapped system into a zero-gap semiconductor, for $\lambda_{\rm ext,R}\approx 1.04$eV, and (c) two pairs of Dirac cones, for $\lambda_{\rm ext,R}\approx 1.2$eV.}
\label{fig:merging}
\end{figure*}
Let us now apply the obtained general results to $\alpha$- and $\gamma$-graphyne.

$\alpha$-graphyne with SOC is effectively described by the same Hamiltonian as graphene [see Eqs.~(\ref{aeff}), (\ref{aiso}), and  (\ref{arashba})]. As a result, for an arbitrary non-zero value of the intrinsic SOC parameter $\lambda_I$ a topological band gap opens [\onlinecite{Kane-Mele-PRL2005}]. The intrinsic SOC opens a non-trivial gap, but respects the spin degeneracy (see Fig.~\ref{fig:alphametiso0,1}). The Rashba SOC leads to trigonal warping and lifts the spin-degeneracy [\onlinecite{sandler,gelderen}] (see also Fig.~\ref{fig:alpharashba}). When both Rashba and intrinsic SOC are present, the spin-degeneracy is lifted, the electron-hole symmetry is broken, and the Rashba SOC tends to close the gap induced by the intrinsic SOC. The main difference compared to graphene is the reduced Fermi velocity and the larger band gap due to the stronger intrinsic SOC in graphynes. Note that the SOC in graphene has been extensively studied by means of first-principles calculations (see Refs.~[\onlinecite{prb80,prb82,prb75,nano9}]).

Whereas $\beta$-graphyne exhibits Dirac cones and the Rashba SOC can destroy the Dirac behavior by opening a gap when the cones merge [\onlinecite{MMJ}], we find that in $\gamma$-graphyne the situation is completely reversed. First of all, we note that in $\gamma$-graphyne the external SOC dominates, simply because by hopping through the acetylene bond one can flip the spin three times as many as compared to hopping through the single bond. The relevant phase diagram is given in Fig.~\ref{fig:2.9}. When the coupling parameter $\lambda_{\rm ext, R}$ is increased, six new pairs emerge along the line connecting $K$ and $K'$  (see region II in Figs.~\ref{fig:2.9} and~\ref{fig:merging}). As the coupling parameter is further increased, we find that six additional pairs of Dirac cones emerge on the line connecting the $\Gamma$ and $M$ points (see region III in Fig.~\ref{fig:2.9}). Merging of the Dirac cones and the corresponding gap opening in shaken honeycomb optical lattices have been recently studied in Ref.\ [\onlinecite{selma}].

To summarize, in this paper we have developed a tight-binding theory for the spin-orbit couplings in graphynes. For completeness, we first considered $\alpha$-, $\beta$-, and $\gamma$-graphyne in absence of the SOC. An effective description in terms of only $p_z$-orbitals captures their band-structures quite well. To describe the SOC, besides the $p_z$-orbitals, we have included the $\sigma-$ and $d-$orbitals, as these two sets of orbitals are now coupled. At half-filling, the latter are away from the Fermi level, and we have integrated them out to obtain an effective TB model in terms of only $p_z$ orbitals at both edges and vertices. In the last step,  the orbitals at the edges have been eliminated and an effective TB model in terms of only orbitals at vertices has been obtained. We have then studied the effective TB models for $\alpha$- and $\gamma$-graphynes, and have repeated some of the results for $\beta$-graphyne, previously published in Ref.\ [\onlinecite{MMJ}], however now in the context of this general theory.

 As a result, the effective low-energy description of $\alpha$-graphyne  differs from graphene with respect to the value of the SOC, and the effective NN-hopping parameter. With respect to $\beta$- and $\gamma$-graphyne, we find that  we have to distinguish between external and internal SOC. In all the three compounds, the effect of the $d$-orbitals is to increase the value of the SOC parameters. We expect that the internal Rashba SOC dominates in $\beta$-graphyne. By tuning an applied electric field, the system can become gapped. On the other hand, for $\gamma$-graphyne  we expect that the external SOC is dominant, and the gap can be closed by applying an electric field. Concerning the intrinsic SOC, we would like to point out that the internal and external SOC can compete with each other. As already shown in Ref.\ [\onlinecite{MMJ}], in $\beta$-graphyne the intrinsic SOC opens a non-trivial gap. On the other hand, in $\gamma$-graphyne the bandgap is topologically trivial, and we estimate that for realistic values of the SOC it cannot be turned into a topological one. We hope that our findings will be a useful base for studying the SOC-related phenomena in other graphynes, such as $6,6,12-$ and $\delta$-graphyne [\onlinecite{JJZheng2013}]. Finally, we also anticipate that our results will motivate further {\it ab initio} studies of the SOCs  in graphynes.

\section{Acknowledgments}
We would like to thank D. S. Galv\~ ao for fruitful discussions. This work is part of the D-ITP consortium, a program of the Netherlands Organization for Scientific Research (NWO) that is funded by the Dutch Ministry of Education, Culture and Science (OCW). The authors acknowledge financial support from NWO and the Dutch FOM association with the program "Designing Dirac carriers in semiconductor honeycomb lattices".

\appendix
\begin{widetext}
\section{Low energy tight-binding Hamiltonian}\label{eff}
Here, we outline the method used in this paper to derive effective Hamiltonians. Consider a system where we can split the spinor $\Psi$ into a high-energy component $\Psi_h$ and a low-energy component $\Psi_l$. Then, we may write the Hamiltonian matrix in a block form
\begin{align}\label{0}
H({\bf k})&=\begin{pmatrix}
H_{ll}({\bf k})&H_{lh}({\bf k})\\
H_{lh}^\dagger({\bf k})&H_{hh}({\bf k})
\end{pmatrix}.
\end{align}
Using this decomposition, the Schr\"odinger equation reads as
\begin{align}
E\Psi_l({\bf k})&=H_{ll}({\bf k})\Psi_l({\bf k})+H_{lh}({\bf k})\Psi_h({\bf k})\label{1}\\
E\Psi_h({\bf k})&=H_{lh}^\dagger({\bf k}) \Psi_l({\bf k})+H_{hh}({\bf k})\Psi_h({\bf k}).\label{2}
\end{align}
 We can then use Eq.~(\ref{2}) to eliminate $\Psi_h({\bf k})$ in Eq.~(\ref{1}). Since $H_{lh}^\dagger({\bf k})\Psi_l({\bf k})=(-H_{hh}({\bf k})+E)\Psi_h({\bf k})$, up to first order in $E$ we obtain $\Psi_h({\bf k})=-H_{hh}^{-1}({\bf k})(1+EH_{hh}^{-1}({\bf k}))H_{lh}^\dagger({\bf k})\Psi_l({\bf k})$. Therefore, Eq.~(\ref{1}) reduces to
\begin{align}\label{4}
(H_{ll}({\bf k})-H_{lh}({\bf k})H_{hh}^{-1}({\bf k})H_{lh}^\dagger({\bf k}))\Psi_l({\bf k})&=E(\mathbb{I}+H_{lh}({\bf k})H_{hh}^{-2}({\bf k})H_{lh}^\dagger({\bf k}))\Psi_l({\bf k}).
\end{align}
If we now introduce $S({\bf k})=\mathbb{I}+H_{lh}({\bf k})H_{hh}^{-2}({\bf k})H_{lh}^\dagger({\bf k})$, and define $\phi({\bf k})=S^{1/2}({\bf k})\Psi_l({\bf k})$, we find the eigenvalue equation
\begin{align}\label{3}
(H_{ll}({\bf k})-H_{lh}({\bf k})H_{hh}^{-1}({\bf k})H_{lh}^\dagger({\bf k}))S^{-1/2}({\bf k})\phi({\bf k})&=E S^{1/2}({\bf k}) \phi({\bf k}).
\end{align}
By multiplying Eq.~(\ref{3}) on both sides with $S^{-1/2}({\bf k})$ we find
\begin{align}
H_{\rm{eff}}({\bf k})\phi({\bf k})&=E\phi({\bf k}),
\end{align}
with $H_{\rm{eff}}({\bf k})$ given by
\begin{align}
H_{\rm{eff}}({\bf k})&=S^{-1/2}({\bf k})(H_{ll}({\bf k})-H_{lh}({\bf k})H_{hh}^{-1}({\bf k})H_{lh}^\dagger({\bf k}))S^{-1/2}({\bf k})
\end{align}
In various cases we consider in this paper the matrix $S$ happens to be diagonal, which greatly simplifies the expressions. It turns out that we can Fourier transform the effective Hamiltonian back to real space, and obtain an effective real space Hamiltonian. Note that several papers adopt a different approach, where they write down the transfer equations, then integrate out the high-energy contributions, see for example Ref.~[\onlinecite{Kim-Choi2012}]. The drawback of their method lies in the fact that this can lead to a non-Hermitian Hamiltonian, as it happens for example in the context of $6,6,12$-graphyne [\onlinecite{zhe}].
\section{Effective Hamiltonian without spin-orbit coupling}\label{appa}
In the main text, we have pointed out that the orbitals located at the vertices correspond to the low-energy states, whereas the orbitals at the edges give rise to high-energy states. This can be seen from the relation $|t_1|<|t_2|<|t_3|$, because the vertices are coupled by the parameter $t_3$, whereas the edges are coupled by the parameter $t_2$. Therefore, we can apply the low-energy approximation to the graphynes we consider, and we are able to accurately describe $\alpha$-graphyne in terms of a two-site model, and $\beta$- and $\gamma$-graphyne by a six-site model. In the  following, we make use of three normalized NN vectors ${\bf d}_i$, $i=1,2,3$ given by
\begin{align}
{\bf d}_1&=(-1/2,\sqrt{3}/2),\nonumber\\
{\bf d}_2&=(1,0)\nonumber\\
{\bf d}_3&=(-1/2,-\sqrt{3}/2).
\end{align}
We denote the different bond lengths by $l_i$, $i=1,2,3$, corresponding to the vertex-vertex, vertex-edge, and edge-edge bonds, respectively.  Furthermore, we define $l_4=2l_2+l_3$.
\subsection{Effective Hamiltonian for $\alpha$-graphyne}\label{appaa}
By performing a Fourier transformation on Eq.~(\ref{eqha}) we obtain
\begin{align}
H^\alpha=\int\mathrm{d}k \Psi^\dagger({\bf k})H({\bf k})\Psi({\bf k}),\nonumber
\end{align}
where $H({\bf k})$ is given by Eq.~(\ref{0}),
\begin{align}
\Psi(k)&=(A({\bf k}),B({\bf k}),a^1({\bf k}),a^2({\bf k}),a^3({\bf k}),b^1({\bf k}),b^2({\bf k}),b^3({\bf k}))^T,
\end{align}
 and  $H_{ll}({\bf k})$, $H_{hl}({\bf k})$, and $H_{hh}({\bf k})$ are given by
\begin{align}
H_{ll}({\bf k})&=0,\nonumber\\
H_{lh}({\bf k})&=t_{\alpha,1}\begin{pmatrix}
e^{i l_2 {\bf k}\cdot {\bf d}_1}& e^{i l_2 {\bf k}\cdot {\bf d}_2}& e^{i l_2 {\bf k}\cdot {\bf d}_3}&0&0&0\\
0&0&0& e^{-i l_2 {\bf k}\cdot {\bf d}_1}& e^{-i l_2 {\bf k}\cdot {\bf d}_2}& e^{-i l_2 {\bf k}\cdot {\bf d}_3}\\
\end{pmatrix},\nonumber\\
H_{hh}({\bf k})&=t_{\alpha,2}\begin{pmatrix}
0& diag(e^{i l_3 {\bf k}\cdot {\bf d}_1},e^{i l_3 {\bf k}\cdot {\bf d}_2},e^{i l_3 {\bf k}\cdot {\bf d}_3})\\
diag(e^{-i l_3 {\bf k}\cdot {\bf d}_1},e^{-i l_3 {\bf k}\cdot {\bf d}_2},e^{-i l_3 {\bf k}\cdot {\bf d}_3})&0
\end{pmatrix}.\nonumber
\end{align}
If we now follow the method outlined in Appendix~A, we obtain
\begin{align}
(H_{ll}({\bf k})-H_{lh}({\bf k})H_{hh}^{-1}({\bf k})H_{lh}^\dagger({\bf k}))&=-\frac{t_{\alpha,1}^2}{t_{\alpha,2}}\begin{pmatrix}0&f({\bf k})\\
f^*({\bf k})&0
\end{pmatrix},\nonumber
\end{align}
with $f({\bf k})=\sum_{j=1}^3 e^{i {\bf k}\cdot {\bf d}_j l_4}$, and $S$ is given by $$S=\mathbb{I}(1+3t_{\alpha,1}^2/t_{\alpha,2}^2)$$. Hence, the effective Hamiltonian reads as
\begin{align}
H_{\rm{eff}}^\alpha({\bf k}) &=-\frac{t_{\alpha,1}^2 t_{\alpha,2}}{t_{\alpha,2}^2+3t_{\alpha,1}^2}
\begin{pmatrix}0&f({\bf k})\\
f^*({\bf k})&0\end{pmatrix},\nonumber
\end{align}
and leads to Eq.~(\ref{aeff}) after performing an inverse Fourier transformation to obtain the real-space Hamiltonian.
\subsection{Effective Hamiltonian for $\beta$-graphyne}\label{appb}
In momentum space the Hamiltonian Eq.~(\ref{eqhb}) reads as
\begin{align}
H^\beta=\int\mathrm{d}{\bf k} \Psi^\dagger({\bf k})H({\bf k})\Psi({\bf k}),\nonumber
\end{align}
where $H({\bf k})$ is given by Eq. (\ref{0}),
\begin{align}
\Psi({\bf k})&=(A({\bf k}),B({\bf k}),C({\bf k}),D({\bf k}),E({\bf k}),F({\bf k}),a^1({\bf k}),b^1({\bf k}),c^1({\bf k}),d^1({\bf k}),e^1({\bf k}),f^1({\bf k}),\nonumber\\
&a^2({\bf k}),b^2({\bf k}),c^2({\bf k}),d^2({\bf k}),e^2({\bf k}),f^2({\bf k}))^T.\nonumber
\end{align}
Here, $H_{ll}({\bf k})$ is given by
\begin{align}
H_{ll}({\bf k})&=t_{\beta,1}\begin{pmatrix}
0&diag(e^{i{\bf k}\cdot {\bf d}_1 l_1},e^{-i{\bf k}\cdot {\bf d}_2 l_1},e^{i{\bf k}\cdot {\bf d}_3 l_1})\\
diag(e^{-i{\bf k}\cdot {\bf d}_1 l_1},e^{i{\bf k}\cdot {\bf d}_2 l_1},e^{-i{\bf k}\cdot {\bf d}_3 l_1})&0
\end{pmatrix}.\nonumber
\end{align}
The matrix $H_{lh}({\bf k})$, which couples electrons at the vertices to the electrons belonging to the acetylene linkage, can be further decomposed as
\begin{align}
H_{lh}({\bf k})&=t_{\beta,2}\begin{pmatrix}H^1_{lh}({\bf k})&H^2_{lh}({\bf k})\\
\end{pmatrix},\nonumber
\end{align}
where $H^1_{lh}({\bf k})$ and $H^2_{lh}({\bf k})$ read
\begin{align}
H^1_{lh}({\bf k})&=diag(e^{i {\bf k}\cdot {\bf d}_3 l_2},e^{-i {\bf k}\cdot {\bf d}_3 l_2},e^{i {\bf k}\cdot {\bf d}_2 l_2},e^{-i {\bf k}\cdot {\bf d}_2 l_2},e^{i {\bf k}\cdot {\bf d}_1 l_2},e^{-i {\bf k}\cdot {\bf d}_1 l_2}),\nonumber\\
H^2_{lh}({\bf k})&=diag(e^{i {\bf k}\cdot {\bf d}_2 l_2},e^{-i {\bf k}\cdot {\bf d}_1 l_2},e^{i {\bf k}\cdot {\bf d}_1 l_2},e^{-i {\bf k}\cdot {\bf d}_3 l_2},e^{i {\bf k}\cdot {\bf d}_3 l_2},e^{-i {\bf k}\cdot {\bf d}_2 l_2}).\nonumber
\end{align}
Finally, we decompose the matrix $H_{hh}({\bf k})$ as
\begin{align}
H_{hh}({\bf k})&=t_{\beta,3}\begin{pmatrix}
H_1({\bf k})&0\\
0&H_2({\bf k})\\
\end{pmatrix}.\nonumber
\end{align}
The matrices $H_1({\bf k})$ and $H_2({\bf k})$ are given by
\begin{align}
H_1({\bf k})&=diag(U_3({\bf k}),U_2({\bf k}),U_1({\bf k})),\nonumber
\end{align}
\begin{align}
H_2({\bf k})&=\begin{pmatrix}
0&0&0&e^{i{\bf k}\cdot {\bf d}_2 l_3}\\
0&U_1(-{\bf k})&0&0\\
0&0&U_3(-{\bf k})&0\\
e^{-i{\bf k}\cdot {\bf d}_2 l_3}&0&0&0
\end{pmatrix},\nonumber
\end{align}
with
\begin{align}
U_i({\bf k})&=\begin{pmatrix}0&e^{i {\bf k}\cdot {\bf d}_i l_3}\\
e^{-i {\bf k}\cdot {\bf d}_i l_3}&0\end{pmatrix}.\nonumber
\end{align}
If we now perform the low-energy approximation, we find that $S=(1+2t_{\beta,2}^2/t_{\beta,3}^2)\mathbb{I}$. Furthermore, if we change the basis to $\tilde\Psi=(A,C,E,B,D,F)$,  then $H_{ll}({\bf k})-H_{lh}({\bf k})H_{hh}^{-1}({\bf k})H_{lh}^\dagger({\bf k})$ reads
\begin{align}
H_{ll}({\bf k})-H_{lh}({\bf k})H_{hh}^{-1}({\bf k})H_{lh}^\dagger({\bf k})&=\begin{pmatrix}
0&U({\bf k})\\
U^\dagger({\bf k})&0
\end{pmatrix},\nonumber
\end{align}
and $U({\bf k})$ reads
\begin{align}
U({\bf k})&=\begin{pmatrix}
\frac{t_{\beta,2}^2}{t_{\beta,3}}e^{i{\bf k}\cdot {\bf d}_3 l_4}&t_{\beta,1}e^{i{\bf k}\cdot {\bf d}_1 l_1}&\frac{t_{\beta,2}^2}{t_{\beta,3}}e^{i{\bf k}\cdot {\bf d}_2 l_4}\\
\frac{t_{\beta,2}^2}{t_{\beta,3}}e^{i{\bf k}\cdot {\bf d}_1 l_4}&\frac{t_{\beta,2}^2}{t_{\beta,3}}e^{i{\bf k}\cdot {\bf d}_2 l_4}&t_{\beta,1}e^{i{\bf k}\cdot {\bf d}_3 l_1}\\
t_{\beta,1}e^{i{\bf k}\cdot {\bf d}_2 l_1}&\frac{t_{\beta,2}^2}{t_{\beta,3}}e^{i{\bf k}\cdot {\bf d}_3 l_4}&\frac{t_{\beta,2}^2}{t_{\beta,3}}e^{i{\bf k}\cdot {\bf d}_1 l_4}
\end{pmatrix}.\nonumber
\end{align}
If we perform an inverse Fourier transformation, we obtain Eq.~(\ref{effb}), the real-space $\beta$-graphyne Hamiltonian.
\subsection{Effective Hamiltonian for $\gamma$-graphyne}\label{appc}
By applying a Fourier transformation on Eq.~(\ref{eqhc}), we obtain
\begin{align}
H^\gamma=\int\mathrm{d}{\bf k} \Psi^\dagger({\bf k})H({\bf k})\Psi({\bf k}),\nonumber
\end{align}
where $H_{\bf k}$ is given by Eq.~(\ref{0}), and
\begin{align}
\Psi({\bf k})&=(A({\bf k}),C({\bf k}),E({\bf k}),B({\bf k}),D({\bf k}),F({\bf k}),a({\bf k}),c({\bf k}),e({\bf k}),b({\bf k}),d({\bf k}),f({\bf k}))^T.\nonumber
\end{align}
The matrix $H_{ll}({\bf k})$ reads
\begin{align}
H_{ll}({\bf k})&=t_{\gamma,1}\begin{pmatrix}0&U_{ll}({\bf k})\\
U_{ll}^\dagger({\bf k})&0
\end{pmatrix},\nonumber
\end{align}
with $U_{ll}({\bf k})$ given by
\begin{align}
U_{ll}({\bf k})&=\begin{pmatrix}
e^{i {\bf k}\cdot {\bf d}_3 l_1}&0&e^{i {\bf k}\cdot {\bf d}_2 l_1}\\
e^{i {\bf k}\cdot {\bf d}_1 l_1}&e^{i {\bf k}\cdot {\bf d}_2 l_1}&0\\
0&e^{i {\bf k}\cdot {\bf d}_3 l_1}&e^{i {\bf k}\cdot {\bf d}_1 l_1}
\end{pmatrix}.\nonumber
\end{align}
Furthermore, we find
\begin{align}
H_{lh}&=t_{\gamma,2}diag(e^{i {\bf k}\cdot {\bf d}_1 l_2},e^{-i {\bf k}\cdot {\bf d}_2 l_2},e^{i {\bf k}\cdot {\bf d}_3 l_2},e^{-i {\bf k}\cdot {\bf d}_1 l_2},e^{i {\bf k}\cdot {\bf d}_2 l_2},e^{-i {\bf k}\cdot {\bf d}_3 l_2}).\nonumber
\end{align}
Finally, $H_{hh}({\bf k})$ reads
\begin{align}
H_{hh}({\bf k})&=t_{\gamma,3}\begin{pmatrix}0&diag(e^{i {\bf k}\cdot {\bf d}_1 l_3},e^{-i {\bf k}\cdot {\bf d}_2 l_3},e^{i {\bf k}\cdot {\bf d}_3 l_3})\\
diag(e^{-i {\bf k}\cdot {\bf d}_1 l_3},e^{i {\bf k}\cdot {\bf d}_2 l_3},e^{-i {\bf k}\cdot {\bf d}_3 l_3})&0
\end{pmatrix},\nonumber
\end{align}
Therefore, we find $S=(1+t_{\gamma,2}^2/t_{\gamma,3}^2)\mathbb{I}$, and
\begin{align}
H_{ll}({\bf k})-H_{lh}({\bf k})H_{hh}^{-1}({\bf k})H_{lh}^\dagger({\bf k})&=\begin{pmatrix}
0&U({\bf k})\\
U^\dagger({\bf k})&0
\end{pmatrix},\nonumber
\end{align}
with
\begin{align}
U({\bf k})&=\begin{pmatrix}
t_{\gamma_1}e^{i{\bf k}\cdot {\bf d}_3 l_4}&\frac{-t_{\gamma,2}^2}{t_{\gamma,3}}e^{i{\bf k}\cdot {\bf d}_1 l_1}&t_{\gamma_1}e^{i{\bf k}\cdot {\bf d}_2 l_4}\\
t_{\gamma_1}e^{i{\bf k}\cdot {\bf d}_1 l_4}&t_{\gamma_1}e^{i{\bf k}\cdot {\bf d}_2 l_4}&\frac{-t_{\gamma,2}^2}{t_{\gamma,3}}e^{i{\bf k}\cdot {\bf d}_3 l_1}\\
\frac{-t_{\gamma,2}^2}{t_{\gamma,3}}e^{i{\bf k}\cdot {\bf d}_2 l_1}&t_{\gamma_1}e^{i{\bf k}\cdot {\bf d}_3 l_4}&t_{\gamma_1}e^{i{\bf k}\cdot {\bf d}_1 l_4}
\end{pmatrix}.\nonumber
\end{align}
As a result, after an inverse Fourier transformation, we end up with the effective Hamiltonian Eq.~(\ref{effc}).
\section{$\sigma$-tight-binding models}\label{sigmatb}
In this appendix we present the TB models containing $\sigma$-orbitals, used to derive the SOC Hamiltonians. To denote different $\sigma$-orbitals on each site, we introduce a subscript $j=1,2,3$. Different sites are labeled by a superscript (the same as for $p_z$-orbitals), see Fig.~\ref{fig:lattices}.
\subsection{$\sigma$-tight-binding model for $\alpha$-graphyne}
The labeling of the different $\sigma$-orbitals is shown in Fig.~\ref{figas}.
 \begin{figure}[t]
\centering
\includegraphics[width=0.8\textwidth]{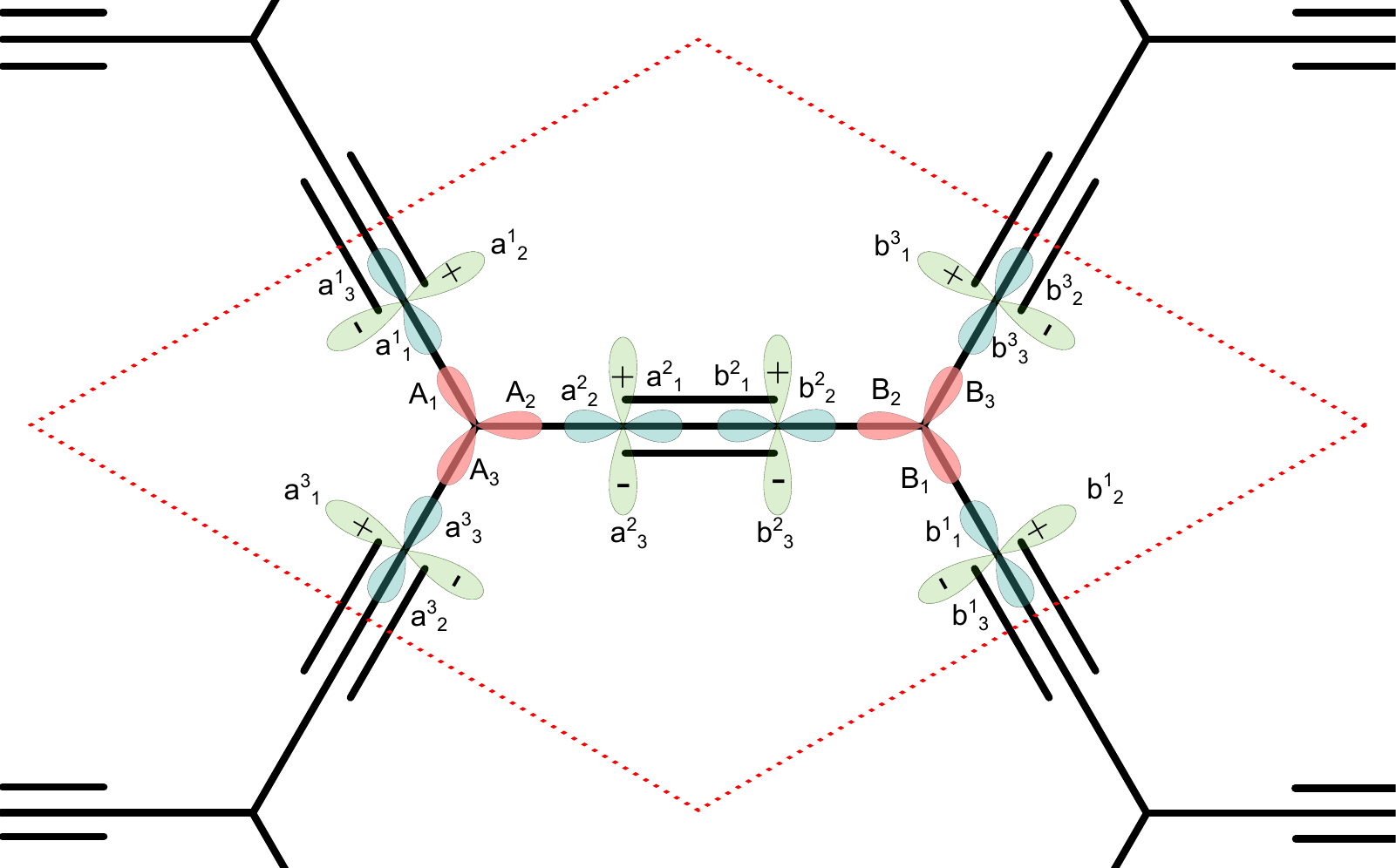}
\caption{(Color online) Labeling used for the $\alpha$-graphyne TB model.}
\label{figas}
\end{figure}
Combining this labeling with the definition of the dominant NN hoppings ($V_2,V_3,V_4$) as given in Fig.~\ref{fig:sigmalattices}(a), we find that $H^\alpha_{NN}$ reads
\begin{align}
H^\alpha_{NN}&=V_2 \sum_{\langle i,j\rangle}\left(A_{1,i}^\dagger a^1_{1,j}+A_{2,i}^\dagger a^2_{2,j}+A_{3,i}^\dagger a^3_{3,j}+B_{1,i}^\dagger b^1_{1,j}+B_{2,i}^\dagger b^2_{2,j}+B_{3,i}^\dagger b^3_{3,j}\right)+V_3 \sum_{\langle i,j\rangle} \left[(a^{1}_{3,i})^\dagger b^{1}_{3,j}+ (a^{2}_{1,i})^\dagger b^{2}_{1,j}\right.\nonumber\\
&\left.+ (a^{3}_{2,i})^\dagger b^{3}_{2,j}\right]+V_4\sum_{\langle i,j\rangle} \left[(a^{1}_{2,i})^\dagger b^{1}_{2,j}+ (a^{2}_{3,i})^\dagger b^{2}_{3,j}+ (a^{3}_{1,i})^\dagger b^{3}_{1,j}\right]+h.c.
\end{align}
The on-site energies ($\varepsilon_2,\varepsilon_3,\varepsilon_4$) and hoppings ($V_6,V_9$) are included in $H^\alpha_{onsite}$, which is given by
\begin{align}
H^\alpha_{onsite}&=\frac{\varepsilon_2}{2}\sum_i\left( A_{1,i}^\dagger A_{1,i}+A_{2,i}^\dagger A_{2,i}+A_{3,i}^\dagger A_{3,i}+ B_{1,i}^\dagger B_{1,i}+B_{2,i}^\dagger B_{2,i}+B_{3,i}^\dagger B_{3,i}\right)\nonumber\\
&+\frac{\varepsilon_3}{2}\sum_i \left[(a^1_{3,i})^\dagger a^1_{3,i}+(a^2_{1,i})^\dagger a^2_{1,i}+(a^3_2)^\dagger a^3_2+(b^1_{3,i})^\dagger b^1_{3,i}+(b^2_{1,i})^\dagger b^2_{1,i}+(b^3_{2,i})^\dagger b^3_{2,i}\right]\nonumber\\
&+\frac{\varepsilon_4}{2}\sum_i \left[(a^1_{1,i})^\dagger a^1_{1,i}+(a^2_{2,i})^\dagger a^2_{2,i}+(a^3_{3,i})^\dagger a^3_{3,i}+(b^1_{1,i})^\dagger b^1_{1,i}+(b^2_{2,i})^\dagger b^2_{2,i}+(b^3_{3,i})^\dagger b^3_{3,i} \right] \nonumber\\
&+V_6\sum_i\left( A_{1,i}^\dagger A_{2,i} +A^\dagger_{2,i} A_{3,i} +A^\dagger_{3,i} A_{1,i}+ B_{1,i}^\dagger B_{2,i} +B^\dagger_{2,i} B_{3,i} +B^\dagger_{3,i} B_{1,i} \right)\nonumber\\
&+V_9 \sum_i \left[(a^1_{1,i})^\dagger a^1_{3,i}+ (a^2_{2,i})^\dagger a^2_{1,i}+ (a^3_{3,i})^\dagger a^3_{2,i}+(b^1_{1,i})^\dagger b^1_{3,i}+ (b^2_{2,i})^\dagger b^2_{1,i}+ (b^3_{3,i})^\dagger b^3_{2,i}\right]+h.c.
\end{align}
\subsection{$\sigma$-tight-binding model for $\beta$-graphyne}\label{c2}
\begin{figure}[t]
\centering
\includegraphics[width=0.5\textwidth]{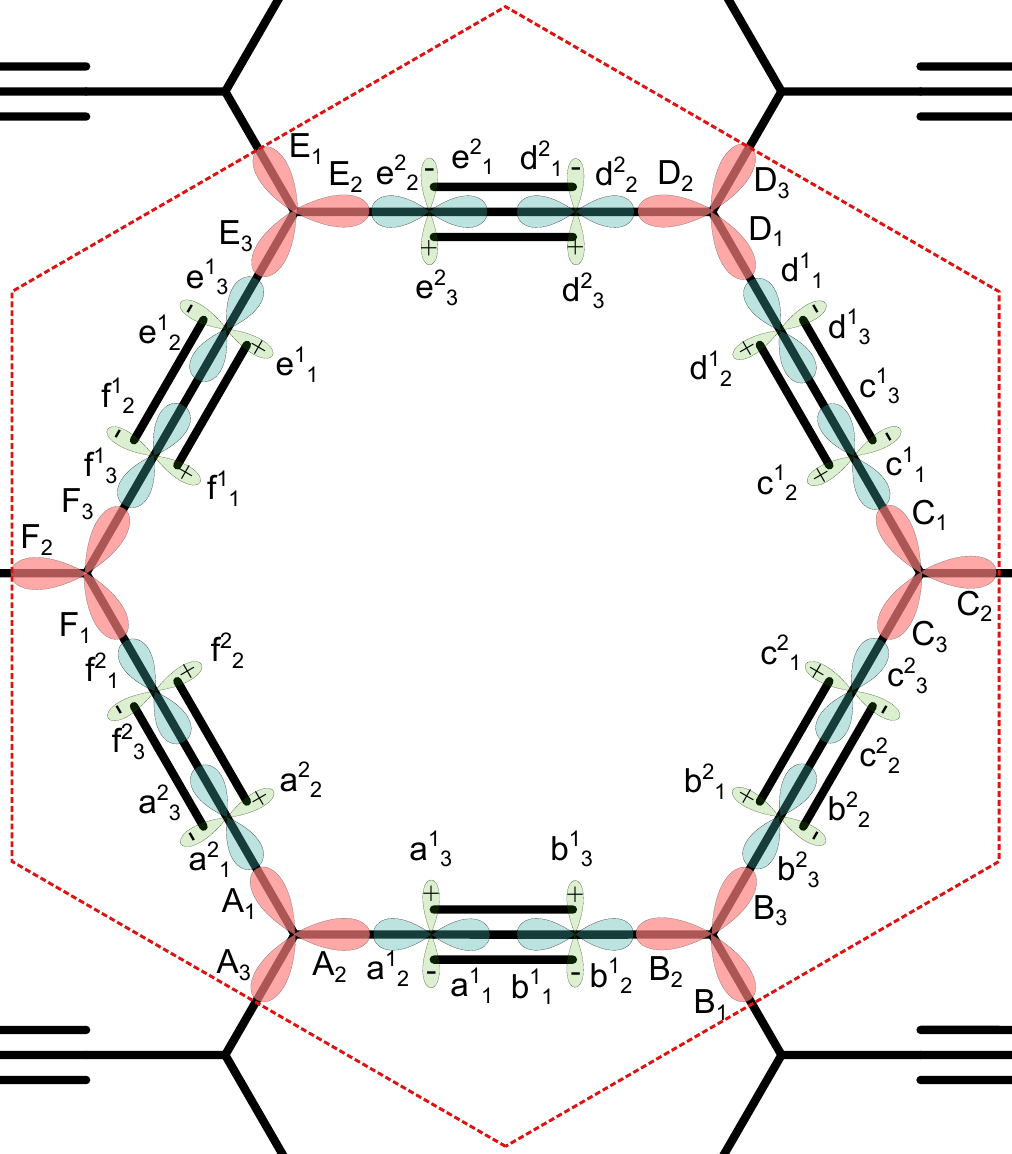}
\caption{(Color online)Labeling of orbitals used in the $\beta$-graphyne TB model.}
\label{figa}
\end{figure}
The labeling of the different $\sigma$-orbitals is shown in Fig.~\ref{figa}, together with the definition of the dominant NN hoppings ($V_1,\ldots,V_4$) shown in Fig.~\ref{fig:sigmalattices}(b). We then find
\begin{align}
H^{\beta}_{{\rm NN}}&=V_1\sum_{\langle i,j\rangle}\left(A_{3,i}^\dagger D_{3,j}+B_{1,i}^\dagger E_{1,j}+C_{2,i}^\dagger F_{2,j}\right)+V_2\sum_{\langle i,j\rangle}\left( A_{1,i}^\dagger a_{1,j}^2+A_{2,i}^\dagger a^1_{2,j}+B_{2,i}^\dagger b_{2,j}^1+B_{3,i}^\dagger b^2_{3,j} +C_{1,i}^\dagger c^1_{1,j}\right.\nonumber\\
&\left.+C_{3,i}^\dagger c_{3,j}^2+D_{1,i}^\dagger d^1_{1,j}+D_{2,i}^\dagger d_{2,j}^2+E_{2,i}^\dagger e^2_{2,j}+E_{3,i}^\dagger e_{3,j}^1+F_{1,i}^\dagger f_{1,j}^2+F_{3,i}^\dagger f_{3,j}^1\right)+V_3\sum_{\langle i,j\rangle} \left[(a^1_{1,i})^\dagger b^1_{1,j}+(a^2_{3,i})^\dagger f^2_{3,j}\right.\nonumber\\
&\left.+(c^1_{3,i})^\dagger d^1_{3,j}+(c^2_{2,i})^\dagger b^2_{2,j}+(e^1_{2,i})^\dagger f^1_{2,j}+(e^2_{1,i})^\dagger d^2_{1,j}\right]+V_4\sum_{\langle i,j\rangle} \left[(a^1_{3,i})^\dagger b^1_{3,j}+(a^2_{2,i})^\dagger f^2_{2,j}+(c^1_{2,i})^\dagger d^1_{2,j}\right.\nonumber\\
&\left.+(c^2_{1,i})^\dagger b^2_{1,j}+(e^1_{1,i})^\dagger f^1_{1,j}+(e^2_{3,i})^\dagger d^2_{3,j}\right]+h.c.,
\end{align}
The term describing the on-site energies ($\varepsilon_1,\ldots,\varepsilon_5$) and hoppings ($V_5,\ldots,V_9$) reads
\begin{align}
H^{\beta}_{{\rm onsite}}&=\frac{\varepsilon_1}{2}\sum_{i}\left( A_{3,i}^\dagger A_{3,i}+B_{1,i}^\dagger B_{1,i}+C_{2,i}^\dagger C_{2,i}+D_{3,i}^\dagger D_{3,i}+E_{1,i}^\dagger E_{1,i}+F_{2,i}^\dagger F_{2,i}\right)+\frac{\varepsilon_2}{2}\sum_{i}\left( A_{1,i}^\dagger A_{1,i}+A_{2,i}^\dagger A_{2,i}\right.\nonumber\\
 &\left.+ B_{2,i}^\dagger B_{2,i}+B_{3,i}^\dagger B_{3,i}+ C_{1,i}^\dagger C_{1,i}+C_{3,i}^\dagger C_{3,i}+D_{1,i}^\dagger D_{1,i}+D_{2,i}^\dagger D_{2,i}+E_{2,i}^\dagger E_{2,i}+E_{3,i}^\dagger E_{3,i}+F_{1,i}^\dagger F_{1,i}+F_{3,i}^\dagger F_{3,i}\right)\nonumber\\
&+\frac{\varepsilon_3}{2}\sum_{i} \left[(a^1_{1,i})^\dagger a^1_{1,i}+(a^2_{3,i})^\dagger a^2_{3,i}+(b^1_{1,i})^\dagger b^1_{1,i}+(b^2_{2,i})^\dagger b^2_{2,i}+(c^1_{3,i})^\dagger c^1_{3,i}+(c^2_{2,i})^\dagger c^2_{2,i}+(d^1_{3,i})^\dagger d^1_{3,i}+(d^2_{1,i})^\dagger d^2_{1,i}\right.\nonumber\\
&\left.+(e^1_{2,i})^\dagger e^1_{2,i}+(e^2_{1,i})^\dagger e^2_{1,i}+(f^1_{2,i})^\dagger f^1_{2,i}+(f^2_{3,i})^\dagger f^2_{3,i}\right]+\frac{\varepsilon_4}{2}\sum_{i}\left[ (a^1_{2,i})^\dagger a^1_{2,i}+(a^2_{1,i})^\dagger a^2_{1,i}+(b^1_{2,i})^\dagger b^1_{2,i}\right.\nonumber\\
&\left.+(b^2_{3,i})^\dagger b^2_{3,i}+(c^1_{1,i})^\dagger c^1_{1,i}+(c^2_{3,i})^\dagger c^2_{3,i}+(d^1_{1,i})^\dagger d^1_{1,i}+(d^2_{2,i})^\dagger d^2_{2,i}+(e^1_{3,i})^\dagger e^1_{3,i}+(e^2_{2,i})^\dagger e^2_{2,i}+(f^1_{3,i})^\dagger f^1_{3,i}+(f^2_{1,i})^\dagger f^2_{1,i}\right]\nonumber\\
&+\frac{\varepsilon_5}{2}\sum_{i} \left[(a^1_{3,i})^\dagger a^1_{3,i}+(a^2_{2,i})^\dagger a^2_{2,i}+(b^1_{3,i})^\dagger b^1_{3,i}+(b^2_{1,i})^\dagger b^2_{1,i}+(c^1_{2,i})^\dagger c^1_{2,i}+(c^2_{1,i})^\dagger c^2_{1,i}+(d^1_{2,i})^\dagger d^1_{2,i}+(d^2_{3,i})^\dagger d^2_{3,i}\right.\nonumber\\
&\left.+(e^1_{1,i})^\dagger e^1_{1,i}+(e^2_{3,i})^\dagger e^2_{3,i}+(f^1_{1,i})^\dagger f^1_{1,i}+(f^2_{2,i})^\dagger f^2_{2,i}\right]+V_5\sum_{i}\left[A_{3,i}^\dagger(A_{1,i}+A_{2,i})+B_{1,i}^\dagger(B_{2,i}+B_{3,i})+C_{2,i}^\dagger(C_{1,i}\right.\nonumber\\
&\left.+C_{3,i})+D_{3,i}^\dagger(D_{1,i}+D_{2,i})+E_{1,i}^\dagger(E_{2,i}+E_{3,i})+F_{2,i}^\dagger(F_{1,i}+F_{3,i})\right]+V_6\sum_i\left[ A_{1,i}^\dagger A_{2,i}+B_{2,i}^\dagger B_{3,i}+C_{1,i}^\dagger C_{3,i}\right.\nonumber\\
&\left.+ D_{1,i}^\dagger D_{2,i}+E_{2,i}^\dagger E_{3,i}+F_{1,i}^\dagger F_{3,i}\right]+V_7\sum_i \left[(a^1_{2,i})^\dagger a^1_{3,i}+(a^2_{1,i})^\dagger a^2_{2,i}+(b^1_{2,i})^\dagger b^1_{3,i}+(b^2_{3,i})^\dagger b^2_{1,i}+(c^1_{1,i})^\dagger c^1_{2,i}\right.\nonumber\\
&\left.+(c^2_{3,i})^\dagger c^2_{1,i}+(d^1_{1,i})^\dagger d^1_{2,i}+(d^2_{2,i})^\dagger d^2_{3,i}+(e^1_{1,i})^\dagger e^1_{3,i}+(e^2_{2,i})^\dagger e^2_{3,i}+(f^1_{3,i})^\dagger f^1_{1,i}+(f^2_{1,i})^\dagger f^2_{2,i}\right]+V_8\sum_i \left[(a^1_{1,i})^\dagger a^1_{3,i}
\right.\nonumber\\
&\left.+(a^2_{3,i})^\dagger a^2_{2,i}+(b^1_{1,i})^\dagger b^1_{3,i}+(b^2_{2,i})^\dagger b^2_{1,i}+(c^1_{3,i})^\dagger c^1_{2,i}+(c^2_{2,i})^\dagger c^2_{1,i}+(d^1_{3,i})^\dagger d^1_{2,i}+(d^2_{1,i})^\dagger d^2_{3,i}+(e^1_{1,i})^\dagger e^1_{2,i}+(e^2_{1,i})^\dagger e^2_{3,i}\right.\nonumber\\
&\left.+(f^1_{2,i})^\dagger f^1_{1,i}+(f^2_{3,i})^\dagger f^2_{2,i}\right]+V_9\sum_i \left[(a^1_{1,i})^\dagger a^1_{2,i}+(a^2_{3,i})^\dagger a^2_{1,i}+(b^1_{1,i})^\dagger b^1_{2,i}+(b^2_{2,i})^\dagger b^2_{3,i}+(c^1_{3,i})^\dagger c^1_{1,i}+(c^2_{2,i})^\dagger c^2_{3,i}\right.\nonumber\\
&\left.+(d^1_{3,i})^\dagger d^1_{1,i}+(d^2_{1,i})^\dagger d^2_{2,i}+(e^1_{3,i})^\dagger e^1_{2,i}+(e^2_{1,i})^\dagger e^2_{2,i}+(f^1_{2,i})^\dagger f^1_{3,i}+(f^2_{3,i})^\dagger f^2_{1,i}\right]+h.c.
\end{align}
\subsection{$\sigma$-tight-binding model for $\gamma$-graphyne}
 \begin{figure}[b]
\centering
\includegraphics[width=0.6\textwidth]{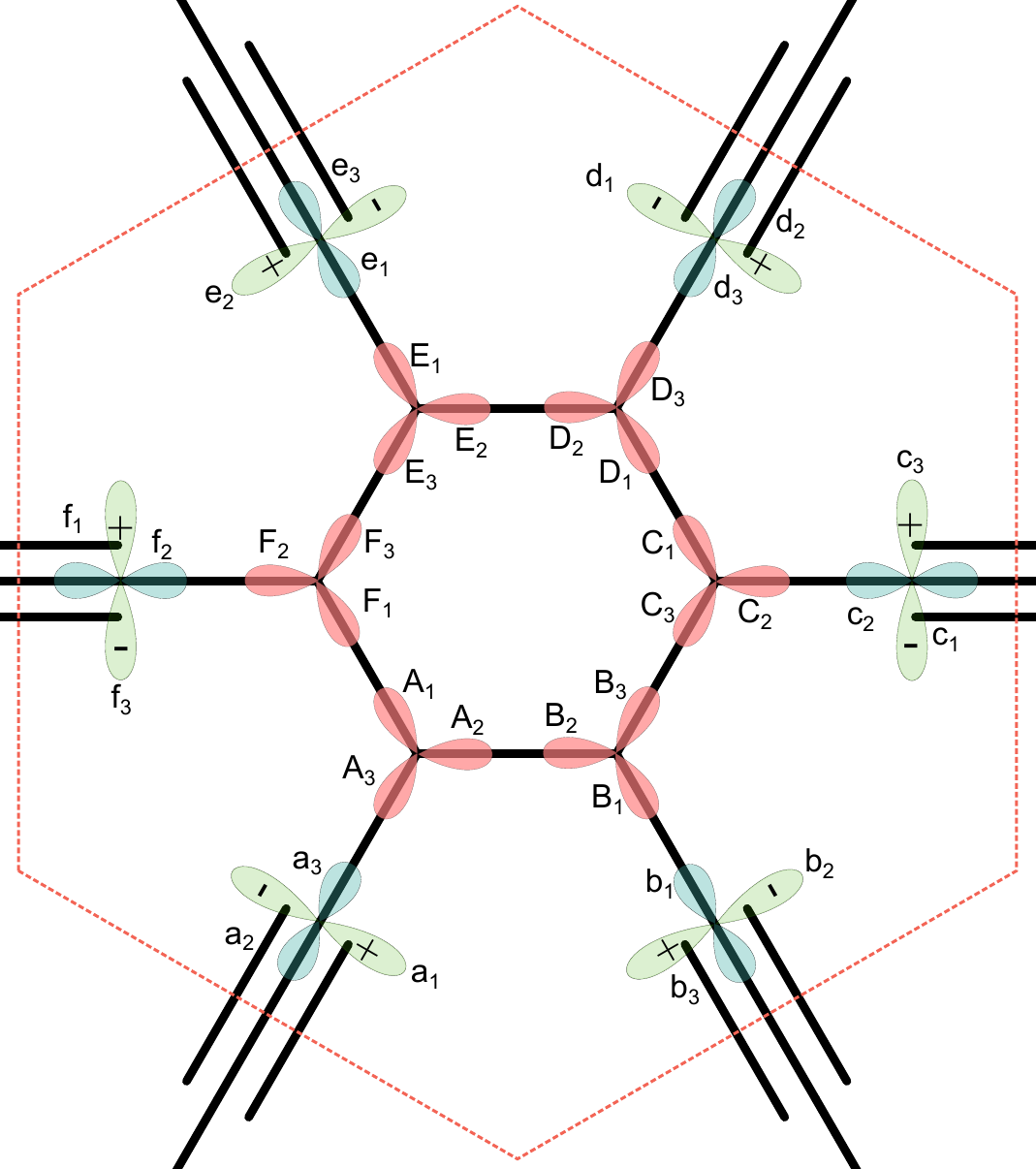}
\caption{(Color online) Labeling used for the $\gamma$-graphyne TB model.}
\label{figag}
\end{figure}
The labeling for $\gamma$-graphyne is shown in Fig.~\ref{figag}. The definition of the dominant NN hoppings ($V_1,\ldots,V_4$) shown in Fig.~\ref{fig:sigmalattices}(c) leads to
\begin{align}
H^\gamma_{NN}&=V_1\sum_{\langle i,j\rangle}\left(A^\dagger_{1,i} F_{1,j}+A^\dagger_{2,i}B_{2,j}+C^\dagger_{1,i}D_{1,j}+C^\dagger_{3,i}B_{3,j}+E^\dagger_{2,i}D_{2,j}+E^\dagger_{3,i}F_{3,j}\right)+V_{2}\sum_{\langle i,j\rangle}\left(A^\dagger_{3,i} a_{3,j}+B^\dagger_{1,i} b_{1,j}\right.\nonumber\\
&\left.+C^\dagger_{2,i} c_{2,j}+D^\dagger_{3,i} d_{3,j}+E^\dagger_{1,i} e_{1,j}+F^\dagger_{2,i} f_{2,j}\right)+V_{3}\sum_{\langle i,j\rangle}\left(a^\dagger_{2,i}d_{2,j}+b^\dagger_{3,i}e_{3,j}+c^\dagger_{1,i} f_{1,j}\right)+V_4\sum_{\langle i,j\rangle}\left(a^\dagger_{1,i}d_{1,j}+b^\dagger_{2,i}e_{2,j}\right.\nonumber\\
&\left.+c^\dagger_{3,i}f_{3,j}\right)+h.c.
\end{align}
The onsite hoppings ($V_5,V_6$) and energies ($\varepsilon_1,\ldots,\varepsilon_5$) yield
\begin{align}
H^\gamma_{onsite}&=\frac{\varepsilon_{1}}{2}\sum_{i}\left( A_{1,i}^\dagger A_{1,i}+A_{2,i}^\dagger A_{2,i} + B_{2,i}^\dagger B_{2,i}+B_{3,i}^\dagger B_{3,i}+ C_{1,i}^\dagger C_{1,i}+C_{3,i}^\dagger C_{3,i}+D_{1,i}^\dagger D_{1,i}+D_{2,i}^\dagger D_{2,i}\right.\nonumber\\
&\left.+E_{2,i}^\dagger E_{2,i}+E_{3,i}^\dagger E_{3,i}+F_{1,i}^\dagger F_{1,i}+F_{3,i}^\dagger F_{3,i}\right)+\frac{\varepsilon_{2}}{2}\sum_{i}\left( A_{3,i}^\dagger A_{3,i}+B_{1,i}^\dagger B_{1,i}+C_{2,i}^\dagger C_{2,i}+D_{3,i}^\dagger D_{3,i}+E_{1,i}^\dagger E_{1,i}\right.\nonumber\\
&\left.+F_{2,i}^\dagger F_{2,i}\right)+\frac{\varepsilon_{3}}{2}\sum_i \left( a^\dagger_{2,i}a_{2,i}+b^\dagger_{3,i}b_{3,i}+c^\dagger_{1,i}c_{1,i}+d^\dagger_{2,i}d_{2,i}+e^\dagger_{3,i}e_{3,i}+f^\dagger_{1,i} f_{1,i}\right)+\frac{\varepsilon_4}{2}\sum_i \left(a^\dagger_{3,i}a_{3,i}+b^\dagger_{1,i}b_{1,i}\right.\nonumber\\
&\left.+c^\dagger_{2,i}c_{2,i}+d^\dagger_{3,i}d_{3,i}+e^\dagger_{1,i}e_{1,i}+f^\dagger_{2,i} f_{2,i}\right)+\frac{\varepsilon_5}{2}\sum_i\left( a^\dagger_{1,i}a_{1,i}+b^\dagger_{2,i}b_{2,i}+c^\dagger_{3,i}c_{3,i}+d^\dagger_{1,i}d_{1,i}+e^\dagger_{2,i}e_{2,i}+f^\dagger_{3,i} f_{3,i}\right)\nonumber\\
&+V_5\sum_i\left[ A_{3,i}^\dagger(A_{1,i}+A_{2,i})+B_{1,i}^\dagger(B_{2,i}+B_{3,i})+C_{2,i}^\dagger(C_{1,i}+C_{3,i})+D_{3,i}^\dagger(D_{1,i}+D_{2,i})+E_{1,i}^\dagger(E_{2,i}+E_{3,i})\right.\nonumber\\
&\left.+F_{2,i}^\dagger(F_{1,i}+F_{3,i})\right]+V_6\sum_i \left(A_{1,i}^\dagger A_{2,i}+B_{2,i}^\dagger B_{3,i}+C_{1,i}^\dagger C_{3,i}+ D_{1,i}^\dagger D_{2,i}+E_{2,i}^\dagger E_{3,i}+F_{1,i}^\dagger F_{3,i}\right)+h.c.
\end{align}
\section{Spin orbit coupling Hamiltonians}
The SOC Hamiltonians $H^{z,\sigma}_E$ and $H^{z,\sigma}_L$ given in Eqs.~(\ref{soch}) and~(\ref{soch2}), respectively, are written in terms of the $p_x$, $p_y$, and $s$ orbitals. Since we would like to compute the  effective SOC Hamiltonians based on Eq.~(\ref{soceff}),  we need to rewrite Eqs.~(\ref{soch}) and~(\ref{soch2}) in terms of the hybrid orbitals. In Table~\ref{basis} we provide the convention we used for the change of basis.
\begin{table}[b]
\resizebox{\linewidth}{!}{%
\begin{tabular}{c||c|c|c|c|c|c|c|c|c|c|c|c|c|c|c|c|c|c}
    &\multicolumn{6}{c|}{$sp^2$}&\multicolumn{6}{c|}{$sp$}&\multicolumn{6}{|c}{$p$}\\\hline
&\includegraphics[scale=0.8]{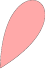} &\includegraphics[scale=0.8]{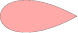}&\includegraphics[scale=0.8]{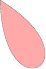}&\includegraphics[scale=0.8]{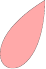}&\includegraphics[scale=0.8]{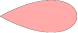}&\includegraphics[scale=0.8]{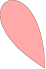}&\includegraphics[scale=0.8]{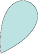}&\includegraphics[scale=0.8]{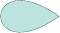}&\includegraphics[scale=0.8]{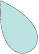}&\includegraphics[scale=0.8]{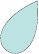} &\includegraphics[scale=0.8]{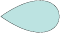}&\includegraphics[scale=0.8]{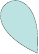}&\includegraphics[scale=0.8]{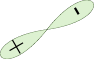}&\includegraphics[scale=0.8]{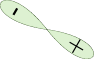}&\includegraphics[scale=0.8]{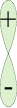}&\includegraphics[scale=0.8]{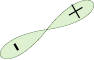}&\includegraphics[scale=0.8]{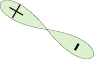}&\includegraphics[scale=0.8]{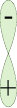}\\\hline
$ s$&$1/\sqrt{3}$&$1/\sqrt{3}$&$1/\sqrt{3}$&$1/\sqrt{3}$&$1/\sqrt{3}$&$1/\sqrt{3}$&$1/\sqrt{2}$&$1/\sqrt{2}$&$1/\sqrt{2}$&$1/\sqrt{2}$&$1/\sqrt{2}$&$1/\sqrt{2}$&0&0&0&0&0&0\\
$ p_x$&$1/\sqrt{6}$&$-\sqrt{2/3}$&$1/\sqrt{6}$&$-1/\sqrt{6}$&$\sqrt{2/3}$&$-1/\sqrt{6}$&$1/\sqrt{8}$&$-1/\sqrt{2}$&$1/\sqrt{8}$&$-1/\sqrt{8}$&$1/\sqrt{2}$&$-1/\sqrt{8}$&$-\sqrt{3}/2$&$\sqrt{3}/2$&$0$&$\sqrt{3}/2$&$-\sqrt{3}/2$&$0$\\
$ p_y$&$1/\sqrt{2}$&$0$&$-1/\sqrt{2}$&$-1/\sqrt{2}$&$0$&$1/\sqrt{2}$&$\sqrt{3/8}$&$0$&$-\sqrt{3/8}$&$-\sqrt{3/8}$&$0$&$\sqrt{3/8}$&$-1/2$&$-1/2$&$1$&$1/2$&$1/2$&$-1$
\end{tabular}}
\caption{(Color online) Overlap between two sets of basis.}
\label{basis}
\end{table}
\subsection{$\alpha$-graphyne}
For $\alpha$-graphyne, the  Hamiltonians~(\ref{soch}) and~(\ref{soch2}) rewritten in terms of the  orbitals read
\begin{align}
H^{z,\sigma}_E&=\xi_{sp2}3^{-1/2}\sum_{i}\left[A_i^\dagger(A_{1,i}+A_{2,i}+ A_{3,i})+B_i^\dagger(B_{1,i}+B_{2,i}+ B_{3,i})\right]+\xi_{sp1}2^{-1/2}\sum_{i}\left[(a^1_i)^\dagger(a^1_{1,i}+a^1_{3,i})+(a^2_i)^\dagger(a^2_{1,i}\right.\nonumber\\
&\left.+a^2_{2,i})+(a^3_i)^\dagger(a^3_{2,i}+a^3_{3,i})+(b^1_i)^\dagger(b^1_{1,i}+b^1_{3,i})+(b^2_i)^\dagger(b^2_{1,i}+b^2_{2,i})+(b^3_i)^\dagger(b^3_{2,i}+b^3_{3,i})\right],\nonumber\\
\end{align}
and
\begin{align}
H^{z,\sigma}_L&=i\xi_{p2}\sum_{i}\left[A^\dagger_i( 2^{-1/2}\sigma_x+6^{-1/2}\sigma_y)A_{1,i}+A^\dagger_i(- (2/3)^{1/2}\sigma_y)A_{2,i}+A^\dagger_i(- 2^{-1/2}\sigma_x+6^{-1/2}\sigma_y)A_{3,i}\right.\nonumber\\
&\left.B^\dagger_i(- 2^{-1/2}\sigma_x-6^{-1/2}\sigma_y)B_{1,i}+B^\dagger_i( (2/3)^{1/2}\sigma_y)B_{2,i}+B^\dagger_i( 2^{-1/2}\sigma_x-6^{-1/2}\sigma_y)B_{3,i}\right]\nonumber\\
&+i\xi_{p1}\sum_{i}\left[(a^1_i)^\dagger2^{-1/2}(-\sqrt{3}\sigma_x/2-\sigma_y/2)a^1_{1,i}+(a^1_i)^\dagger(\sigma_x/2-\sqrt{3}\sigma_y/2)a^1_{2,i}+(a^1_i)^\dagger2^{-1/2}(\sqrt{3}\sigma_x/+\sigma_y/2)a^1_{3,i}\right.\nonumber\\
&+(a^2_i)^\dagger(-2^{-1/2}\sigma_y)a^2_{1,i}+(a^2_i)^\dagger(2^{-1/2}\sigma_y)a^2_{2,i}+(a^2_i)^\dagger(\sigma_x)a^2_{3,i}\nonumber\\
&+(a^3_i)^\dagger(\sigma_x/2+\sqrt{3}\sigma_y/2)a^3_{1,i}+(a^3_i)^\dagger2^{-1/2}(-\sqrt{3}\sigma_x/2+\sigma_y/2)a^3_{2,i}+(a^3_i)^\dagger2^{-1/2}(\sqrt{3}\sigma_x/2-\sigma_y/2)a^3_{3,i}\nonumber\\
&+(b^1_i)^\dagger2^{-1/2}(\sqrt{3}\sigma_x/2+\sigma_y/2)b^1_{1,i}+(b^1_i)^\dagger(\sigma_x/2-\sqrt{3}\sigma_y/2)b^1_{2,i}+(b^1_i)^\dagger2^{-1/2}(-\sqrt{3}\sigma_x/2-\sigma_y/2)b^1_{3,i}\nonumber\\
&+(b^2_i)^\dagger2^{-1/2}(\sigma_y)b^2_{1,i}-(b^2_i)^\dagger2^{-1/2}(\sigma_y)b^2_{2,i}+(b^2_i)^\dagger(\sigma_x)b^2_{3,i}\nonumber\\
&\left.+(b^3_i)^\dagger(\sigma_x/2+\sqrt{3}\sigma_y/2)b^3_{1,i}+(b^3_i)^\dagger2^{-1/2}(\sqrt{3}\sigma_x/2-\sigma_y/2)b^3_{2,i}+(b^3_i)^\dagger2^{-1/2}(-\sqrt{3}\sigma_x/2+\sigma_y/2)b^3_{3,i}\right]
\end{align}
\subsection{$\beta$-graphyne}
Similarly, for $\beta$-graphyne we find
\begin{align}
H^{z,\sigma}_{E}&=\xi_{sp2}3^{-1/2}\sum_i\left[ A_i^\dagger(A_{1,i}+A_{2,i}+A_{3,i})+B_i^\dagger(B_{1,i}+B_{2,i}+B_{3,i})+C_i^\dagger(C_{1,i}+C_{2,i}+C_{3,i})+D_i^\dagger(D_{1,i}+D_{2,i}\right.\nonumber\\
&\left.+D_{3,i})+E_i^\dagger(E_{1,i}+E_{2,i}+E_{3,i})+F_i^\dagger(F_{1,i}+F_{2,i}+F_{3,i})\right]+\xi_{sp1}2^{-1/2}\sum_i\left[ (a^1_{i})^\dagger(a^1_{2,i}+a^1_{1,i})+(a^2_{i})^\dagger(a^2_{1,i}+a^2_{3,i})\right.\nonumber\\
&+(b^1_{i})^\dagger(b^1_{2,i}+b^1_{1,i})+(b^2_{i})^\dagger(b^2_{2,i}+b^2_{3,i})+(c^1_{i})^\dagger(c^1_{1,i}+c^1_{3,i})+(c^2_{i})^\dagger(c^2_{2,i}+c^2_{3,i})+(d^1_{i})^\dagger(d^1_{1,i}+d^1_{3,i})\nonumber\\
&\left.+(d^2_{i})^\dagger(d^2_{1,i}+d^2_{2,i})+(e^1_{i})^\dagger(e^1_{3,i}+e^1_{2,i})+(e^2_{i})^\dagger(e^2_{1,i}+e^2_{2,i})+(f^1_{i})^\dagger(f^1_{2,i}+f^1_{3,i})+(f^2_{i})^\dagger(f^2_{1,i}+f^2_{3,i})\right],
\end{align}
and
\begin{align}
H_{L}^{z,\sigma}&=i\xi_{p2}\sum_{i}\left[A^\dagger_{i}( 2^{-1/2}\sigma_x+6^{-1/2}\sigma_y)A_{1,i}+A^\dagger_{i}(- (2/3)^{1/2}\sigma_y)A_{2,i}+A^\dagger_{i}(- 2^{-1/2}\sigma_x+6^{-1/2}\sigma_y)A_{3,i}\right.\nonumber\\
&+B^\dagger_{i}(-2^{-1/2}\sigma_x-6^{-1/2}\sigma_y)B_{1,i}+B^\dagger_{i}( (2/3)^{1/2}\sigma_y)B_{2,i}+B^\dagger_{i}( 2^{-1/2}\sigma_x-6^{-1/2}\sigma_y)B_{3,i}\nonumber\\
&+C^\dagger_{i}( 2^{-1/2}\sigma_x+6^{-1/2}\sigma_y)C_{1,i}+C^\dagger_{i}(-(2/3)^{1/2}\sigma_y)C_{2,i}+C^\dagger_{i}(-2^{-1/2}\sigma_x+6^{-1/2}\sigma_y)C_{3,i}\nonumber\\
&+D^\dagger_{i}(- 2^{-1/2}\sigma_x-6^{-1/2}\sigma_y)D_{1,i}+D^\dagger_{i}( (2/3)^{1/2}\sigma_y)D_{2,i}+D^\dagger_{i}( 2^{-1/2}\sigma_x-6^{-1/2}\sigma_y)D_{3,i}\nonumber\\
&+E^\dagger_{i}( 2^{-1/2}\sigma_x+6^{-1/2}\sigma_y)E_{1,i}+E^\dagger_{i}(-(2/3)^{1/2}\sigma_y)E_{2,i}+E^\dagger_{i}(-2^{-1/2}\sigma_x+6^{-1/2}\sigma_y)E_{3,i}\nonumber\\
&\left.+F^\dagger_{i}(- 2^{-1/2}\sigma_x-6^{-1/2}\sigma_y)F_{1,i}+F^\dagger_{i}( (2/3)^{1/2}\sigma_y)F_{2,i}+F^\dagger_{i}( 2^{-1/2}\sigma_x-6^{-1/2}\sigma_y)F_{3,i}\right]\nonumber\\
&+i\xi_{p1}\sum_i\left[(a^1_{i})^\dagger(-2^{-1/2}\sigma_y)a^1_{1,i}+(a^1_{i})^\dagger(2^{-1/2}\sigma_y)a^1_{2,i}+(a^1_{i})^\dagger(\sigma_x)a^1_{3,i}+(a^2_{i})^\dagger2^{-1/2}(-\sqrt{3}\sigma_x/2-\sigma_y/2)a^2_{1,i}\right.\nonumber\\
&+(a^2_{i})^\dagger(\sigma_x/2-\sqrt{3}\sigma_y/2)a^2_{2,i}+(a^2_{i})^\dagger2^{-1/2}(\sqrt{3}\sigma_x/2+\sigma_y/2)a^2_{3,i}+(b^1_{i})^\dagger2^{-1/2}\sigma_y b^1_{1,i}-(b^1_{i})^\dagger2^{-1/2}\sigma_y b^1_{2,i}\nonumber\\
&+(b^1_{i})^\dagger\sigma_x b^1_{3,i}+(b^2_{i})^\dagger (\sigma_x/2+\sqrt{3}\sigma_y/2)b^2_{1,i}+(b^2_{i})^\dagger 2^{-1/2}(\sqrt{3}\sigma_x/2-\sigma_y/2)b^2_{2,i}-(b^2_{i})^\dagger 2^{-1/2}(\sqrt{3}\sigma_x/2-\sigma_y/2)b^2_{3,i}\nonumber\\
&+(c^2_{i})^\dagger(\sigma_x/2+\sqrt{3}\sigma_y/2)c^2_{1,i}-(c^2_{i})^\dagger 2^{-1/2}(\sqrt{3}\sigma_x/2-\sigma_y/2)c^2_{2,i}+(c^2_{i})^\dagger 2^{-1/2}(\sqrt{3}\sigma_x/2-\sigma_y/2)c^2_{3,i}\nonumber\\
&+(c^1_{i})^\dagger2^{-1/2}(-\sqrt{3}\sigma_x/2-\sigma_y/2)c^1_{1,i}-(c^1_{i})^\dagger(\sigma_x/2-\sqrt{3}\sigma_y/2)c^1_{2,i}+(c^1_{i})^\dagger2^{-1/2}(\sqrt{3}\sigma_x/2+\sigma_y/2)c^1_{3,i}\nonumber\\
&-(d^1_{i})^\dagger2^{-1/2}(-\sqrt{3}\sigma_x/2-\sigma_y/2)d^1_{1,i}-(d^1_{i})^\dagger(\sigma_x/2-\sqrt{3}\sigma_y/2)d^1_{2,i}-(d^1_{i})^\dagger2^{-1/2}(\sqrt{3}\sigma_x/2+\sigma_y/2)d^1_{3,i}\nonumber\\
&-(d^2_{i})^\dagger(-2^{-1/2}\sigma_y)d^2_{1,i}-(d^2_{i})^\dagger(2^{-1/2}\sigma_y)d^2_{2,i}-(d^2_{i})^\dagger(\sigma_x)d^2_{3,i}\nonumber\\
&-(e^1_{i})^\dagger (\sigma_x/2+\sqrt{3}\sigma_y/2)e^1_{1,i}-(e^1_{i})^\dagger 2^{-1/2}(\sqrt{3}\sigma_x/2-\sigma_y/2)e^1_{2,i}+(e^1_{i})^\dagger 2^{-1/2}(\sqrt{3}\sigma_x/2-\sigma_y/2)e^1_{3,i}\nonumber\\
&-(e^2_{i})^\dagger2^{-1/2}\sigma_y e^2_{1,i}+(e^2_{i})^\dagger2^{-1/2}\sigma_y e^2_{2,i}-(e^2_{i})^\dagger\sigma_x e^2_{3,i}\nonumber\\
&-(f^1_{i})^\dagger (\sigma_x/2+\sqrt{3}\sigma_y/2)f^1_{1,i}+(f^1_{i})^\dagger 2^{-1/2}(\sqrt{3}\sigma_x/2-\sigma_y/2)f^1_{2,i}-(f^1_{i})^\dagger 2^{-1/2}(\sqrt{3}\sigma_x/2-\sigma_y/2)f^1_{3,i}\nonumber\\
&\left.(f^2_{i})^\dagger2^{-1/2}(\sqrt{3}\sigma_x/2+\sigma_y/2)f^2_{1,i}+(f^2_{i})^\dagger(\sigma_x/2-\sqrt{3}\sigma_y/2)f^2_{2,i}-(f^2_{i})^\dagger2^{-1/2}(\sqrt{3}\sigma_x/2+\sigma_y/2)f^2_{3,i}\right].
\end{align}
\subsection{$\gamma$-graphyne}
The SOC Hamiltonians for $\gamma$-graphyne are given by
\begin{align}
H^{z,\sigma}_E&=\xi_{sp2}3^{-1/2}\sum_i\left[ A_i^\dagger(A_{1,i}+A_{2,i}+A_{3,i})+B_i^\dagger(B_{1,i}+B_{2,i}+B_{3,i})+C_i^\dagger(C_{1,i}+C_{2,i}+C_{3,i})+D_i^\dagger(D_{1,i}+D_{2,i}\right.\nonumber\\
&\left.+D_{3,i})+E_i^\dagger(E_{1,i}+E_{2,i}+E_{3,i})+F_i^\dagger(F_{1,i}+F_{2,i}+F_{3,i})\right]+\xi_{sp1}2^{-1/2}\sum_i\left[a_i^\dagger(a_{2,i}+a_{3,i})+b_i^\dagger(b_{1,i}+b_{3,i})\right.\nonumber\\
&\left.c_i^\dagger(c_{1,i}+c_{2,i})+d_i(d_{2,i}+d_{3,i})+e_i(e_{1,i}+e_{3,i})+f_i(f_{1,i}+f_{2,i})\right],
\end{align}
and
\begin{align}
H_{L}^{z,\sigma}&=i\xi_{p2}\sum_{i}\left[A^\dagger_{i}( 2^{-1/2}\sigma_x+6^{-1/2}\sigma_y)A_{1,i}+A^\dagger_{i}(- (2/3)^{1/2}\sigma_y)A_{2,i}+A^\dagger_{i}(- 2^{-1/2}\sigma_x+6^{-1/2}\sigma_y)A_{3,i}\right.\nonumber\\
&+B^\dagger_{i}(-2^{-1/2}\sigma_x-6^{-1/2}\sigma_y)B_{1,i}+B^\dagger_{i}( (2/3)^{1/2}\sigma_y)B_{2,i}+B^\dagger_{i}( 2^{-1/2}\sigma_x-6^{-1/2}\sigma_y)B_{3,i}\nonumber\\
&+C^\dagger_{i}( 2^{-1/2}\sigma_x+6^{-1/2}\sigma_y)C_{1,i}+C^\dagger_{i}(-(2/3)^{1/2}\sigma_y)C_{2,i}+C^\dagger_{i}(-2^{-1/2}\sigma_x+6^{-1/2}\sigma_y)C_{3,i}\nonumber\\
&+D^\dagger_{i}(- 2^{-1/2}\sigma_x-6^{-1/2}\sigma_y)D_{1,i}+D^\dagger_{i}( (2/3)^{1/2}\sigma_y)D_{2,i}+D^\dagger_{i}( 2^{-1/2}\sigma_x-6^{-1/2}\sigma_y)D_{3,i}\nonumber\\
&+E^\dagger_{i}( 2^{-1/2}\sigma_x+6^{-1/2}\sigma_y)E_{1,i}+E^\dagger_{i}(-(2/3)^{1/2}\sigma_y)E_{2,i}+E^\dagger_{i}(-2^{-1/2}\sigma_x+6^{-1/2}\sigma_y)E_{3,i}\nonumber\\
&\left.+F^\dagger_{i}(- 2^{-1/2}\sigma_x-6^{-1/2}\sigma_y)F_{1,i}+F^\dagger_{i}( (2/3)^{1/2}\sigma_y)F_{2,i}+F^\dagger_{i}( 2^{-1/2}\sigma_x-6^{-1/2}\sigma_y)F_{3,i}\right]\nonumber\\
&+i\xi_{p1}\sum_i \left[ a^\dagger_i(-\sigma_x/2-\sqrt{3}\sigma_y/2)a_{1,i}+a^\dagger_i2^{-1/2}(-\sqrt{3}\sigma_x/2+\sigma_y/2)a_{2,i}+a^\dagger_i2^{-1/2}(\sqrt{3}\sigma_x/2-\sigma_y/2)a_{3,i}\right.\nonumber\\
&+b^\dagger_i2^{-1/2}(\sqrt{3}\sigma_x/2+\sigma_y/2)b_{1,i}+b^\dagger_i(-\sigma_x/2+\sqrt{3}\sigma_y/2)b_{2,i}+b^\dagger_i2^{-1/2}(-\sqrt{3}\sigma_x/2-\sigma_y/2)b_{3,i}\nonumber\\
&-c^\dagger_i2^{-1/2}\sigma_y c_{1,i}+c^\dagger_i2^{-1/2}\sigma_y c_{2,i}+c^\dagger_i\sigma_x c_{3,i}\nonumber\\
&+d^\dagger_i(-\sigma_x/2-\sqrt{3}\sigma_y/2)d_{1,i}+d^\dagger_i2^{-1/2}(\sqrt{3}\sigma_x/2-\sigma_y/2)d_{2,i}+d^\dagger_i2^{-1/2}(-\sqrt{3}\sigma_x/2+\sigma_y/2)d_{3,i}\nonumber\\
&+e^\dagger_i2^{-1/2}(-\sqrt{3}\sigma_x/2-\sigma_y/2)e_{1,i}+e^\dagger_i(-\sigma_x/2+\sqrt{3}\sigma_y/2)e_{2,i}+e^\dagger_i2^{-1/2}(\sqrt{3}\sigma_x/2+\sigma_y/2)e_{3,i}\nonumber\\
&\left.+f^\dagger_i2^{-1/2}\sigma_y f_{1,i}-f^\dagger_i2^{-1/2}\sigma_y f_{2,i}+f^\dagger_i\sigma_x f_{3,i}\right]
\end{align}
\end{widetext}

\end{document}